\documentclass[11pt]{article}
\date{}

\usepackage[latin1]{inputenc}
\usepackage{epsfig}
\usepackage{color}
\usepackage{graphics}
\usepackage{graphicx}

\usepackage{amsthm}
\usepackage{amssymb}
\usepackage{amsmath}
\usepackage{amsfonts}

\numberwithin{equation}{section}

\begin{document}

\newcommand{\df}{\stackrel{\rm def}{=}}
\newcommand{\co}{{\scriptstyle \circ}}
\newcommand{\lb}{\lbrack}
\newcommand{\rb}{\rbrack}
\newcommand{\rn}[1]{\romannumeral #1}
\newcommand{\msc}[1]{\mbox{\scriptsize #1}}
\newcommand{\dsp}{\displaystyle}
\newcommand{\scs}[1]{{\scriptstyle #1}}

\newcommand{\ket}[1]{| #1 \rangle}
\newcommand{\bra}[1]{| #1 \langle}
\newcommand{\vac}{| \mbox{vac} \rangle }

\newcommand{\e}{\mbox{{\bf e}}}
\newcommand{\va}{\mbox{{\bf a}}}
\newcommand{\bc}{\mbox{{\bf C}}}

\newcommand{\com}{C\!\!\!\!|}

\newcommand{\br}{\mbox{{\bf R}}}
\newcommand{\bz}{\mbox{{\bf Z}}}
\newcommand{\bq}{\mbox{{\bf Q}}}
\newcommand{\bn}{\mbox{{\bf N}}}
\newcommand {\eqn}[1]{(\ref{#1})}

\newcommand{\cp}{\mbox{{\bf P}}^1}
\newcommand{\n}{\mbox{{\bf n}}}
\newcommand{\sbz}{\msc{{\bf Z}}}
\newcommand{\sn}{\msc{{\bf n}}}

\newcommand{\be}{\begin{equation}}\newcommand{\ee}{\end{equation}}
\newcommand{\bea}{\begin{eqnarray}} \newcommand{\eea}{\end{eqnarray}}
\newcommand{\ba}[1]{\begin{array}{#1}} \newcommand{\ea}{\end{array}}

\newcommand{\cleqn}{\setcounter{equation}{0}}

\makeatletter

\@addtoreset{equation}{section}

\def\theequation{\thesection.\arabic{equation}}
\makeatother

\def\oa{\bigcirc\!\!\!\! a}
\def\ob{\bigcirc\!\!\!\! b}
\def\oc{\bigcirc\!\!\!\! c}
\def\oi{\bigcirc\!\!\!\! i}
\def\oj{\bigcirc\!\!\!\! j}
\def\ok{\bigcirc\!\!\!\! k}
\def\ve{\vec e}\def\vk{\vec k}\def\vn{\vec n}\def\vp{\vec p}
\def\vr{\vec r}\def\vs{\vec s}\def\vt{\vec t}\def\vu{\vec u}
\def\vv{\vec v}\def\vx{\vec x}\def\vy{\vec y}\def\vz{\vec z}

\title{Fusion rules and four-point functions\\
in
the AdS$_3$ WZNW model}

\author{Walter H. Baron  $^{1,}$ \footnote{e-mail: w\_baron@iafe.uba.ar} \, and
  Carmen~A.~N\'u\~nez $^{1,~2,}$
\footnote{e-mail: carmen@iafe.uba.ar}$^\dag$}

\date{\small $^1$ Instituto de Astronom\'{\i}a y F\'{\i}sica del Espacio
  (CONICET-UBA).\\
C.~C.~67 - Suc.~28, 1428 Buenos Aires, Argentina \\ and \\
  $^2$ Departamento de F\'\i sica,
FCEN, Universidad de Buenos Aires. \\ Ciudad Universitaria, Pab. I,
 1428 Buenos Aires,
Argentina.}

\maketitle

\begin{abstract}

We study the operator product expansion in
the 
AdS$_3$ WZNW model. The OPE of primary fields and their spectral flow
images is computed from
 the analytic continuation of
the expressions in the H$_3^+$ WZNW model, adding spectral flow.
We argue that the symmetries of the affine algebra require a
truncation 
which establishes the closure of the fusion rules on the Hilbert space
of the theory.
Although the physical mechanism determining the decoupling
is not completely understood, 
we present several consistency checks on the results. A preliminary  analysis of 
 factorization 
allows to obtain some properties of four-point functions involving fields in
generic sectors of the theory,
to verify that they agree with the spectral flow
selection rules and to show
that the truncation must be realized in
physical amplitudes for consistency.

\end{abstract}

\newpage

\tableofcontents

\section{Introduction}

String theory on AdS$_3$ with NS antisymmetric background field
is one of the best understood string theories
in curved geometries and it
has been very useful for the analysis
of black holes in two and three dimensions and of some cosmological
spacetimes. It  is so far the only case in which the AdS/CFT
correspondence \cite{malda} can be
examined beyond the supergravity approximation with  control
over the worldsheet theory, and this property allowed to show, in particular,
 the 
equivalence among three-point correlators of BPS observables in
the superstring on AdS$_3\times$ S$^3\times$ T$^4$ and those of the dual
conformal field theory (CFT) \cite{all}.

The worldsheet of the bosonic string propagating on AdS$_3$
is described by the
$SL(2, \mathbb R)$ WZNW model. The string
spectrum
 is built from affine
primaries of a product of left and right copies of
the universal cover of $SL(2, \mathbb R)$ and their spectral flow images
\cite{mo1}. It consists of
 long strings with continuous energy spectrum
arising from the principal
 continuous representation and its spectral
flow images, and short strings with discrete physical spectrum
resulting from the highest-weight
discrete representation and its spectral flow images.
A
no ghost theorem for this spectrum was proved in \cite{mo1} and
verified in \cite{mo2}.
Amplitudes  on the sphere were computed in \cite{mo3},
analytically continuing the expressions obtained for the  Euclidean
H$^+_3=\frac{SL(2,\mathbb C)}{SU(2)}$ WZNW model in \cite{tesch1,
  tesch2}. 
 Some subtleties of the analytic continuation relating the H$^+_3$
and AdS$_3$ models were clarified in \cite{mo3} and this
allowed to construct, in particular,
the  four-point function of unflowed  short
strings.
Integrating over
the moduli space of the worldsheet,
it was shown that the  string amplitude
can be expressed as a sum of products
of three-point functions with intermediate physical
states, $i.e.$ the structure of the factorization agrees with the Hilbert
space of the theory.

A step up towards a proof of consistency and unitarity of the theory
involves the
construction of  four-point functions including states 
in different representations and the
verification
that
only unitary states corresponding to long and short strings in agreement with the
spectral flow selection rules  are produced in the
intermediate channels. 
To achieve this goal,
the analytic and algebraic structure of
the  $SL(2,\mathbb R)$ WZNW model should be explored
further.

 Most of the important progress achieved in \cite{mo3} is based
on the better though not yet completely  understood Euclidean H$_3^+$
model. Together with Liouville theory, these are examples of
non-rational conformal field theories with continuous families of
primary fields. The absence of singular vectors and the lack of chiral
factorization
in the relevant current
algebra representations obstruct
the use of the powerful techniques from
rational conformal field theories.
Nevertheless, a  generalized conformal bootstrap approach was successfully
applied in \cite{tesch1, tesch2}  to the
H$_3^+$ model on the punctured sphere, allowing to discuss the
factorization of four-point functions.  In principle, this method
offers the possibility to
unambiguously determine any
$n>3-$point function
in terms of two- and three-point functions once
the operator product expansions
of two operators and the structure
constants are known.
Aiming  to carry out some initial steps
towards developing this procedure for the more involved
AdS$_3$ WZNW model, in these notes
we examine
the role of the spectral flow symmetry on the analytic
continuation of the operator product expansion from H$_3^+$ to  the
relevant representations of $SL(2,
\mathbb R)$ and on the factorization  properties of four-point functions.

While only contributions of highest-weight states are usually written
in an OPE, the descendants being neglected, a
 fundamental problem of the AdS$_3$ WZNW model is that the
spectral flow operation maps primaries into descendants and viceversa.
Thus, to complete this programme it is necessary to learn more about
 the spectral flow representations and
the
secondary fields than is currently known. 
Nevertheless, based on
previous work in \cite{mo3}-\cite{ribault},
we are
able to make some progress. We obtain the OPE of
fields in all sectors of the theory and discuss some properties of the
factorization limit of four-point
functions.

The paper is organized as follows. In section 2 we review some well-known
results on the H$_3^+$ and  AdS$_3$ WZNW models
 in order to setup the notations. In section 3 we 
 analytically continue the expressions obtained in
\cite{tesch1,tesch2} from the Euclidean to the
Lorentzian model and we add spectral flow to obtain the
OPE of primary fields and their spectral flow images.
 The extension of the OPE to  generic descendants is discussed in
 section 4 where
  we show that the spectral flow symmetry requires a
 truncation of the fusion rules determining the closure of the
 operator algebra on the Hilbert space of the theory. In
 section 5 we consider the
factorization of  four-point
functions and study some of its properties.
Finally, section 6 contains a summary and  conclusions. Some technical
 details of the
calculations are included in  appendices A.1 and A.3 and the relation of our
results to certain conclusions in \cite{mo3} is the content of appendix A.2.

\section{Review of the H$_3^+$ and AdS$_3$ WZNW models}

In this section we review some
well-known results on the H$_3^+$ and the AdS$_3$ WZNW
models
 in order to setup the notations.

A thorough study of the H$_3^+=\frac {SL(2,\mathbb C)}{SU(2)}$
WZNW model was presented in
 \cite{tesch1, tesch2}. The 
Lagrangian formulation was  developed in \cite{gawe} and it
follows from
\begin{equation}
{\cal
  L}=k(\partial\phi\overline\partial\phi+e^{2\phi}\overline\partial\gamma
\partial\overline\gamma)\, .
\end{equation}
Normalizable operators
$\Phi_j(x,\overline x; z,\overline z)$, $x,z \in \mathbb C$, are
labeled by the spin $j=-\frac 12 +i{\mathbb R}_+$ of a principal
continuous  representation of $SL(2,\mathbb C)$ and can be
semiclassically identified with the  expression
\begin{equation}
\Phi_j(x,\overline x; z,\overline z)
=\frac{2j+1}{\pi}\left ((\gamma-x)(\overline\gamma-\overline
x)e^\phi+e^{-\phi}
\right )^{2j}\, . \label{phij}
\end{equation}
They
satisfy
the following OPE
with the holomorphic $SL(2,\mathbb C)$ currents
\begin{equation}
J^a(z)\Phi_j(x,\overline x;z',\overline
z')\sim\frac{D^a\Phi_j(x,\overline x;z',\overline z')}{z-z'}\, ,\qquad a=\pm, 3\, ,
\end{equation}
where
$D^-=\partial_x\, ,  D^3=x\partial_x-j\, ,
D^+=x^2\partial_x-2jx$, and they have conformal weight
 $\widetilde\Delta=-\frac{j(j+1)}{k-2}$.
The asymptotic  $\phi\rightarrow\infty$ expansion, given by
\begin{equation}
\Phi_j(x,\overline x| z,\overline z)
\sim :e^{2(-1-j)\phi(z)}:\delta^2\left (\gamma(z)-x\right
)+B(j):e^{2j\phi(z)}: |\gamma(z)-x|^{4j}\, ,\label{asym}
\end{equation}
 fixes a normalization and determines the relation
between
$\Phi_j$ and $\Phi_{-1-j}$ as
\begin{equation}
\Phi_j(x,\overline x| z,\overline z)= B(j)
\int_{\mathbb C}d^2x'|x-x'|^{4j}\Phi_{-1-j}(x',\overline x';
z,\overline z)\, ,
\end{equation}
where the reflection coefficient $B(j)$ is given by
\begin{eqnarray}
B(j) =   \frac{k-2}{\pi} \,
\frac{\nu^{1+2j}}{\gamma\left (-\frac{1+2j}{k-2}\right )}\, , \quad
\nu=\pi\frac{\Gamma\left (1-\frac 1{k-2}\right )}{\Gamma\left (1+\frac
  1{k-2}\right )}\, ,\quad \gamma(x)=\frac{\Gamma(x)}{\Gamma(1-x)} \, .
\label{2pf}
\end{eqnarray}
For our purposes, it is convenient to transform the primary fields to
the $m-$basis as
\begin{equation}
\Phi^j_{m,\overline m}(z,\overline z)=\displaystyle\int d^2x ~ x^{j+m}
~\overline
x^{j+\overline m}~\Phi_{-1-j}(x,\overline x; z,\overline z)\, ,
\label{phim}
\end{equation}
where $m=\frac {n+is}2,~ \overline m=\frac{-n+is}2,~
n\in {\mathbb Z}, s\in \mathbb R$. The fields $\Phi^j_{m,\overline m}$
have the following OPE with the chiral
currents
\begin{equation}
J^\pm(z)\Phi^j_{m,\overline m}(z',\overline z')\sim\frac{\mp j +
  m}{z-z'}
\Phi^j_{m\pm 1,\overline
m}(z',\overline z')\, ,\quad J^3(z)\Phi^j_{m,\overline m}(z',\overline z')
\sim\frac m{z-z'}\Phi^j_{m,\overline m}(z',\overline z')\, ,
\end{equation}
and the relation between
$\Phi^j_{m,\overline m}$ and $\Phi^{-1-j}_{m,\overline m}$ is given
by 
\begin{equation}
\Phi^j_{m,\overline m}(z,\overline z)=B(-1-j) c^{-1-j}_{m,\overline
  m}\Phi^{-1-j}_{m,\overline m}(z,\overline z)
 = \frac{\pi B(-1-j)}{\gamma(2+2j)}\frac{\Gamma(1+j+m)\Gamma(1+j-\overline m)}
{\Gamma(-j+m)\Gamma(-j-\overline m)}\Phi^{-1-j}_{m,\overline
  m}(z,\overline z)\, .\label{gnm}
\end{equation}

The following operator product expansion for any product $\Phi_{j_1}\Phi_{j_2}$
was determined in \cite{tesch1,
  tesch2}:
\begin{eqnarray}
\Phi_{j_2}(x_2|z_2)\Phi_{j_1}(x_1|z_1)&=&\displaystyle\int_{{\cal P}^+}dj_3~
  C(-j_1,-j_2,-j_3)~|z_2-z_1|^{-\widetilde\Delta_{12}}\int_{\mathbb
  C}d^2x_3|x_1-x_2|^{2j_{12}}\nonumber\\
&&\times
|x_1-x_3|^{2j_{13}}
  |x_2-x_3|^{2j_{23}}
\Phi_{-1-j_3}(x_3|z_1)+{\rm descendants}.\label{ope}
\end{eqnarray}
Here, the integration contour is ${\cal P}^+=-\frac
  12+i{\mathbb R}_+$,
the structure constants $C(j_i)$ are given by
\begin{eqnarray}
C(j_1,j_2,j_3) = - \frac{G(1- j_1-j_2-j_3)G(-j_{12})G(-j_{13})G(-j_{23})}{2 \pi^2
  \nu^{j_1+j_2+j_3-1} \gamma\left(\frac{k-1}{k-2}\right) G(-1)
  G(1-2j_1) G(1-2j_2)  G(1-2j_3)}, \label{sc}
\end{eqnarray}
with
$G(j)=(k-2)^{\frac{j(1-j-k)}{2(k-2)}} \, \Gamma_2(-j |1,k-2 ) \, \Gamma_2(k-1+j |1,k-2 ),
$
$\Gamma_2(x|1,w)$ being the Barnes double Gamma function,
$\widetilde\Delta_{12}=\widetilde\Delta(j_1)+\widetilde\Delta(j_2)-
\widetilde\Delta(j_3)$ and $j_{12}=j_1+j_2-j_3$, etc.

The OPE (\ref{ope}) holds
for a
range of values of $j_1, j_2$ given by
\begin{equation}
|{\rm Re}(j_{21}^\pm)|<\frac 12\, ,\qquad j_{21}^+=j_2+j_1+1\, ,\qquad
 j_{21}^-=j_2-j_1\, .\label{range}
\end{equation}
This is the maximal region in which $j_1, j_2$ may vary such that
none of the poles of the integrand hits the contour of integration
over $j_3$. However, as long as the imaginary parts of $j_{21}^\pm$
do not vanish, J. Teschner \cite{tesch2} showed that (\ref{ope})
admits an analytic continuation to generic complex values of $j_1,
j_2$, defined by deforming the contour ${\cal P}^+$. The deformed
contour is given by the sum of the original one plus a finite number
of circles around the poles leading to a finite sum of residue
contributions to the OPE. When $j_{21}^\pm$ are real one can give
them a small imaginary part which is sent to zero after deforming
the contour.

Inserting (\ref{ope}) into a  four-point function
gives an expansion of the correlator which takes the form of an
integral with respect
to the spin of the intermediate representation. The integrand
factorizes into structure constants, two-point functions and conformal blocks.
Since these expressions are analytic in $j$'s (up to delta functions),
correlation functions involving states with arbitrary spin values
may be  obtained through an
appropriate analytic continuation. This procedure was implemented in
\cite{mo3}
to construct the  four-point function of short strings in AdS$_3$.

The worldsheet of the string propagating  on AdS$_3$ is described by
the $SL(2,\mathbb R)$ WZNW model which shares the $\widehat
{sl(2)}$ symmetries with the H$_3^+$ model though it
differs in the allowed representations.
The spectrum of the AdS$_3$ WZNW model
was determined in
\cite{mo1} and it
is constructed from
a product of
left and right copies of representations  of the universal cover of $SL(2,\mathbb R)$.
It is built on
 products of conventional representations of the zero modes,
$i.e.$ the
 principal continuous
representations ${\cal C}_j^\alpha\otimes {\cal C}_j^\alpha$
with
$j=-\frac 12 + i{\mathbb R}, \alpha = (0,1]$ and
the 
lowest-weight discrete series
${\cal D}^+_j\otimes{\cal D}^+_j$ with $j\in\mathbb R$ and $-\frac{k-1}2<j<-\frac
12$. It contains
the current algebra descendants $\widehat
{\cal C}_j^\alpha\otimes \widehat{\cal C}_j^\alpha$,
$\widehat
{\cal D}^+_j\otimes\widehat{\cal D}^+_j$,
and  spectral flow images $\widehat{\cal C}_j^{\alpha,w}\otimes
\widehat{\cal C}_j^{\alpha,w}$, $\widehat
{\cal D}^{+,w}_j\otimes\widehat{\cal D}^{+,w}_j$,
with the same value of $j$ and the same
amount of spectral flow on the
left and
right sectors. Throughout this paper we deal with these representations
of the universal cover of
$SL(2, \mathbb R)$, to which we refer as $SL(2,\mathbb R)$ for short.

The spectral flow representations are generated by the following
 automorphism of the current algebra
\begin{equation}
\tilde J_n^3   =   J_n^3-\frac k2 w\delta_{n,0}\, ,\qquad
\tilde J^\pm_n   =  J^\pm_{n\pm w}\,\label{flowmap1} ,
\end{equation}
with $w\in \mathbb Z$, which gives a copy of the
Virasoro algebra with
\begin{equation}
\tilde L_n = L_n + wJ^3_n-\frac k4 w^2\,\label{flowmap2} .
\end{equation}
Unlike in the compact $SU(2)$ case, different amounts of spectral flow give inequivalent
representations  
of the current algebra of $SL(2,\mathbb R)$.

An affine primary state in the unflowed sector is mapped by the
 automorphism (\ref{flowmap1}) to a
highest/lowest-weight state of the global $sl(2)$ algebra.
We denote these fields in the spectral flow
sector $w$ as
$\Phi^{j,w}_{m,\overline m}$. Their explicit expressions will not be
 needed below.
It is only necessary to know that they verify the following OPE with the currents:
\begin{equation}
J^3(z)\Phi^{j,w}_{m,\overline m}(z',\overline z')\sim\frac{m+\frac
  k2w}{z-z'}\Phi^{j,w}_{m,\overline m}(z',\overline z')\, ,\quad
  J^\pm(z)\Phi^{j,w}_{m,\overline m}(z',\overline z')\sim\frac{\mp j+m}{(z-z')^{\pm
  w}}\Phi^{j,w}_{m\pm 1,\overline m}(z',\overline z')+\cdots\, \nonumber
\end{equation}
and
$m-\overline m \in \mathbb Z$, $m+\overline m\in \mathbb R$.

Two- and three-point functions of the fields $\Phi_j(x|z)$ in the H$_3^+$ model
were computed in \cite{tesch1, tesch2}.
Following \cite{mo3, hs, satoh, gk} we assume that  correlation functions
of primary fields in the
$SL(2,\mathbb R)$ WZNW
model are those of H$_3^+$ with
$j_i, m_i,
\overline m_i$ taking values in representations of
$SL(2,\mathbb R)$.
The  spectral flow operation
 is straightforwardly performed in the
$m-$basis where the only change in the $w-$conserving expectation values of
 fields $\Phi_{m,\overline m}^{j,w}$ in different $w$ sectors is in the
 powers of the coordinates $z_i, \overline z_i$. 
Correlation functions may violate $w-$conservation according to
 the following spectral flow
selection rules
\bea
-N_t+2 ~ \le ~ \sum_{i=1}^{N_t} &w_i&  \le ~ N_c-2\, ,\qquad {\rm {at
    ~least~ one ~
state ~in}}~
\widehat{\cal C}^{\alpha, w}_j
\otimes\widehat{\cal C}^{\alpha, w}_j , \label{cc}\\
-N_d+1~\le ~ \sum_{i=1}^{N_t} &w_i & \le ~ -1\, ,\qquad\qquad {\rm {all~
    states ~in}}~
\widehat{\cal D}_j^{+,w}
\otimes\widehat{\cal D}_j^{+,w},\label{dd}
\eea
with $N_t=N_c+N_d$ and $N_c, N_d$ are the total numbers of operators in
$\widehat{\cal C}^{\alpha, w}_j\otimes\widehat{\cal C}^{\alpha, w}_j
$ and $\widehat{\cal D}^{+,w}_j\otimes\widehat{\cal D}^{+,w}_j$,
respectively.

The spectral flow preserving two-point function is given by
\begin{eqnarray}
\label{2point}
 \langle \Phi^{j,w}_{m,\overline m}(z,\overline z)\Phi^{j',-w}_{m',\overline
  m'}(z', \overline z')\rangle &=&  \delta^2(m+m') \,
  (z-z')^{-2\Delta(j)}
  (\overline{z}-\overline{z}')^{-2\overline{\Delta}(j)}  \nonumber \\
&&
\times ~ \left[ \delta(j+j'+1) +  B(-1-j) c^{-1-j}_{m,\overline{m}} \,
  \delta(j-j')
\right], \label{218}
\end{eqnarray}
where
$\Delta(j) =
\widetilde\Delta(j)-wm-\frac{k}{4} w^2 =
-\frac{j(j+1)}{k-2}-wm-\frac{k}{4} w^2$. For
 states in discrete series it is convenient to work with 
spectral flow images of both
 lowest- and highest-weight representations
related by the
 identification 
$\widehat{\cal D}_j^{+, w}\equiv \widehat{\cal
  D}_{-\frac k2-j}^{-, w+ 1}$, which determines the range of values
 for the spin
\begin{equation}
-\frac{k-1}2<j<-\frac 12\, , \label{unitary}
\end{equation}
and allows to obtain the ($\pm 1$) unit spectral flow
two-point functions from (\ref{218}).

Spectral flow conserving three-point functions
are the following:
\begin{eqnarray}
\label{threepoint}
&& \left\langle \prod_{i=1}^3 \Phi^{j_i,w_i}_{m_i,\overline
  m_i}(z_i,\overline z_i)
\right\rangle = \delta^2({\mbox{$\sum m_i$}}) \, C(1+j_i)
W\left[\begin{matrix}
j_1\,,\,j_2\,,\,j_3\cr m_1,m_2,m_3\cr\end{matrix}\right] \prod_{i<j}
z_{ij}^{-\Delta_{ij}}
\overline z_{ij}^{-\overline \Delta_{ij}} ,
\end{eqnarray}
where $z_{ij}=z_i-z_j$ and $C(j_i)$ is given by
(\ref{sc}). The function $W$
is 
\begin{eqnarray}
&& W\left[\begin{matrix}
j_1\,,\,j_2\,,\,j_3\cr m_1,m_2,m_3\cr\end{matrix}\right] =
\int d^2x_1\, d^2x_2\, x_1^{j_1+m_1}
\overline{x}_1^{j_1+\overline{m}_1} x_2^{j_2+m_2}
\overline{x}_2^{j_2+\overline{m}_2}
 \nonumber \\
&& ~~~~~~~~~~~~~~~~~~~~ \qquad\qquad
\times ~ |1-x_1|^{-2j_{13}-2} |1-x_2|^{-2j_{23}-2}
|x_1-x_2|^{-2j_{12}-2}\, ,\label{w1i}
\end{eqnarray}
and we omit the obvious
$\overline m-$dependence in the arguments to lighten the notation.
This integral was computed in \cite{fukuda}.

The one unit spectral flow
 three-point function  \cite{mo3}
is given by \footnote{For
 an independent calculation of three-point functions using the free
 field approach
see \cite{in}}
\begin{eqnarray}
\label{3pointwind}
 \left\langle \prod_{i=1}^3 \Phi^{j_i,w_i}_{m_i,\overline{m}_i}(z_i,
 \overline z_i)
\right\rangle =  \delta^2(\mbox{$\sum m_i$}\pm\frac k2) \,
\frac{\widetilde{C}(1+j_i)
  \widetilde{W}\left[\begin{matrix}
j_1~~,~j_2~,~~j_3\cr \pm m_1,\pm m_2,\pm m_3\cr\end{matrix}\right]}
{\gamma(j_1+j_2+j_3+3-\frac k2)}
\, \prod_{i<j} z_{ij}^{-\Delta_{ij}}
\overline{z}_{ij}^{-\overline{\Delta}_{ij}} ,
\end{eqnarray}
where
$\sum_iw_i=\pm 1$, the $\pm$ signs corresponding to the
$\pm$ signs in the r.h.s.,
\begin{eqnarray}
\widetilde{C}(j_i) \sim B(-j_1) C\left (\frac k2-j_1,j_2,j_3 \right ),
\end{eqnarray}
up to $k-$dependent, $j-$independent factors and
\begin{eqnarray}
  \widetilde{W}\left[\begin{matrix}
j_1\,,\,j_2\,,\,j_3\cr  m_1, m_2, m_3\cr\end{matrix}\right]=\frac{\Gamma(1+j_1+m_1)}
{\Gamma(-j_1-\overline{m}_1)}
\frac{\Gamma(1+j_2+\overline m_2)}{\Gamma(-j_2-m_2)}
\frac{\Gamma(1+j_3+\overline{m}_3)} {\Gamma(-j_3-m_3)}.
\end{eqnarray}

For discrete states, this expression is related
  to the $\sum_iw_i=\pm 2$
 three-point function through $\widehat{\cal D}_j^{+, w}\equiv \widehat{\cal
  D}_{-\frac k2-j}^{-, w+ 1}$.

In the following sections we shall use these results to study the analytic
continuation of the OPE (\ref{ope}) from the H$_3^+$ 
to the AdS$_3$ WZNW model.
Then, we shall discuss some aspects of the factorization of
four-point functions.

\section{Operator algebra in the ${SL(2,\mathbb R)}$ WZNW model}

A non-trivial check on
the   OPE (\ref{ope})
and structure constants (\ref{sc}) of
the H$_3^+$ WZNW model is that 
the well-known
fusion rules of degenerate representations \cite{ay} 
are exactly recovered by analytically continuing  $j_i, i=1,2$  \cite{tesch1}. 
 On the other hand, it
was argued in \cite{mo3, tesch1, tesch2, hs,satoh, gk} that correlation
functions  in the H$_3^+$ and AdS$_3$ WZNW models are related
by analytic continuation and moreover, the
$k\rightarrow \infty$ limit of the
OPE of  unflowed fields computed along these lines in
\cite{hs, satoh} exhibits complete agreement with the classical tensor
products of representations of $SL(2,\mathbb R)$ \cite{holman}.
It seems then natural to conjecture that the 
 OPE of all fields in the spectrum of the AdS$_3$ WZNW
model can be obtained from
(\ref{ope}) 
analytically continuing $j_1, j_2$ from the range
 $(\ref{range})$.

 However, the 
 spectral flowed fields do not belong to the spectrum of the H$_3^+$
 model and moreover,
the spectral flow symmetry
 transforms primaries into descendants. Thus,
 a better knowledge of these representations seems necessary in
 order to obtain  the fusion rules in the AdS$_3$ model. Nevertheless,
we will show that it is possible to obtain them from the H$_3^+$ model by analytic
 continuation and  by
taking into account 
the $w-$violating structure constants in addition to (\ref{sc}).
 In this section we 
explore this possibility in order to get the OPE of primary fields and their
spectral flow images in the $SL(2,\mathbb R)$ WZNW model.

To deal with highest/lowest-weight and spectral flow representations
it is convenient to work in the $m-$basis.  We have to keep in mind
that when $j$ is real, new divergences appear in the transformation
from the $x-$basis and it must be performed for certain values of
$m_i,\overline m_i$, $i=1,2$. Indeed, to transform the OPE (\ref{ope}) 
to the $m-$basis using (\ref{phim}), the integrals over
$x_1, x_2$ in the r.h.s. must be interchanged with the integral over $j_3$ and
 this process does not commute in general if
there are divergences. However, restricting $j_1, j_2$ to the
range (\ref{range}), 
one can check that the
integrals commute and are regular when  $|m_i|<\frac 12$ and
$|\overline m_i|<\frac12$, $i=1,2,3$, where $m_3=m_1+m_2, \overline
m_3=\overline m_1+\overline m_2$. For other values of $m_i,
\overline m_i$ the OPE 
must be defined, as usual, by analytic continuation of the
parameters. 
Therefore, after performing the $x_1, x_2$ integrals, the
OPE 
(\ref{ope}) in
the $m-$basis is found to be
\begin{eqnarray}
\left.\Phi^{j_1}_{m_1,\overline m_1}(z_1, \overline z_1)\Phi^{j_2}_{m_2,\overline
  m_2}(z_2, \overline z_2)\right|_{w=0}&=&
\displaystyle\int_{\cal P} dj_3~
|z_{12}|^{-2\widetilde\Delta_{12}}~Q^{w=0}\left[
\begin{matrix}
j_1\,,j_2,\,j_3\cr  m_1, m_2, m_3\cr\end{matrix}\right]
\Phi^{j_3}_{m_3,\overline
m_3}(z_1,\overline z_2)\cr
&&~~~~~~~+~{\rm descendants} ,\label{opetm}
\end{eqnarray}
where 
we have defined
\begin{equation}
Q^{w=0}\left[
\begin{matrix}
j_1\,,j_2,\,j_3\cr  m_1, m_2, m_3\cr\end{matrix}\right]=C(1+j_1,1+j_2,-j_3)W
\left[
\begin{matrix}
j_1,j_2,-1-j_3\cr  m_1, m_2, -m_3\cr\end{matrix}\right]\, .\label{q0}
\end{equation}

It is easy to see that the integrand is symmetric under
$j_3\rightarrow-1-j_3$ using the identity \cite{satoh}
\begin{eqnarray}
\frac{W\left[\begin{matrix}
j_1\,,\,j_2\,,\,j_3\cr  m_1, m_2, m_3\cr\end{matrix}\right]}{W\left[\begin{matrix}
j_1,j_2,-1-j_3\cr  m_1\,,\, m_2\,,\, m_3\cr\end{matrix}\right]}
=\frac{C(1+j_1,1+j_2,-j_3)}
{C(1+j_1,1+j_2,1+j_3)}B(-1-j_3)c^{-1-j_3}_{m_3,\overline m_3}\, ,\label{eqsat}
\end{eqnarray}
and as a consequence of (\ref{gnm}). 
In the $x-$basis, every pole in (\ref{ope}) appears duplicated, one over
 the real axis
and another one below, and
the 
 $j_3\rightarrow-1-j_3$ symmetry implies that the integral may be equivalently
 performed either over ${\rm Im} ~j_3>0$ or
over ${\rm Im}~ j_3<0$  \cite{tesch2}.
 In the
 $m-$basis, the $(j_1,j_2)-$dependent poles are also duplicated but
the $m-$dependent poles  are not. The 
 $j_3\rightarrow-1-j_3$ symmetry is still
present, as we discussed above, because of poles and zeros in the
normalization of $\Phi^j_{m,\overline m}$.
The integral must be extended to the full axis
${\cal P}=-\frac12+i\mathbb R$ before
performing the analytic continuation in $m_1,\, m_2$ because the $m-$dependent
poles fall on the real axis.
The maximal regions in
 which $m_1,m_2$ may vary such that
 none of the
poles hit the contour of integration are
$|m_1+m_2|<-\frac
 12$ and $|\overline m_1+\overline m_2|<-\frac 12$.

Since
  the $w-$conserving structure constants of  operators $\Phi_{m,\overline
  m}^{j,w}\in {\cal C}_{j}^{\alpha, w}$ or ${\cal
  D}_{j}^{+,w}$ in
  different $w$ sectors do not change in the $m-$basis 
\footnote{~ We denote the series 
containing the highest/lowest-weight states
 obtained by spectral flowing primaries as ${\cal C}_{j}^{\alpha, w},{\cal
  D}_{j}^{+,w}$.},
the OPE (\ref{opetm}) should also hold  for fields 
obtained by spectral flowing primaries to arbitrary $w$
  sectors, as long as they satisfy $w_1+w_2=w_3$.
 But this OPE would yield an incorrect zero answer 
if  used to compute a $w-$violating three-point function.
It seems then natural to  additionally take into account the spectral
  flow non-preserving structure constants and consider the
  following OPE \footnote{A similar expression was proposed in
 \cite{ribault} and some supporting evidence
was presented from
the relation between the H$_3^+$ model and Liouville theory.}
\begin{eqnarray}
\Phi^{j_1,w_1}_{m_1,\overline m_1}(z_1,\overline z_1)\Phi^{j_2,w_2}_{m_2,\overline
  m_2}(z_2, \overline z_2)
= \sum_{w=-1}^1
\displaystyle\int_{{\cal P}} dj_3~Q^w
 z_{12}^{-\Delta_{12}}\overline
  z_{12}^{-\overline\Delta_{12}}
\Phi^{j_3,w_3}_{m_3,\overline
m_3}(z_2,\overline z_2) +\cdots\, , \label{opec}
\end{eqnarray}
with $w=w_3-w_1-w_2$,
 $m_3=m_1+m_2-\frac k2w$, $\overline m_3=\overline
m_1+\overline
m_2-\frac k2w$, and

\begin{eqnarray}
Q^{w=\pm 1}(j_i;m_i,\overline m_i)  &=&
\widetilde W\left[\begin{matrix}
j_1\,,\,j_2\,,\,j_3\cr  \mp m_1, \mp m_2,
\pm m_3\cr\end{matrix}\right]\frac{\widetilde C(j_i+1)}
{B(-1-j_3)
c_{m_3,\overline m_3}^{-j_3-1}\gamma(j_1+j_2+j_3+3-\frac{k}{2})}\nonumber\\
&\sim &
\frac{\Gamma(\pm\overline m_3-j_3)}{\Gamma(1+j_3\mp m_3)}
\prod_{a=1}^2\frac{\Gamma(1+j_a\mp m_a)}
{\Gamma(-j_a\pm\overline m_a)}
\frac{C(\frac{k}{2}-1-j_1,1+j_2,1+j_3)}
{\gamma(j_1+j_2+j_3+3-\frac{k}{2})}
\, .\nonumber\\
\label{qw}
\end{eqnarray}
For completeness, according to the spectral flow selection rules (\ref{dd}),
 we should also 
include terms with $w=\pm 2$ in the sum. However, we shall show
 in the next section that these do not affect
 the results of the OPE.
The integrand is symmetric under $j_3\rightarrow -1-j_3$. This follows
from (\ref{eqsat}) and the analogous identity
\begin{eqnarray}
\frac{\widetilde {W}\left[\begin{matrix}
j_1\,,\,j_2\,,\,j_3\cr  m_1, m_2,
m_3\cr\end{matrix}\right]}{\widetilde {W}\left[\begin{matrix}
j_1,j_2,-1-j_3\cr  m_1\,,\, m_2\,,\, m_3\cr\end{matrix}\right]}
=\frac{\widetilde {C}(1+j_1,1+j_2,-j_3)\gamma(j_1+j_2+j_3+3-\frac k2)}
{\widetilde {C}(1+j_1,1+j_2,1+j_3)\gamma(j_1+j_2-j_3+2-\frac k2)}B(-1-j_3)c^{-1-j_3}_{m_3,\overline
  m_3}\, ,
\label{eqsat2}
\end{eqnarray}
together with the reflection relation
\begin{equation}
\Phi^{j,w}_{m,\overline m}(z,\overline z)=B(-1-j) c^{-1-j}_{m,\overline
  m}\Phi^{-1-j,w}_{m,\overline m}(z,\overline z)\, . \label{rr}
\end{equation}
The dots in (\ref{opec}) stand for spectral flow images
of current
algebra descendants with the
same $J_0^3$ eigenvalues $m_3,\overline m_3$.
This expression is
 valid for $j_1, j_2$ in the range (\ref{range}) and
the restrictions on
$m_1,m_2$   depend on $Q^w$. The maximal regions in
 which they
 may vary such that
 none of the
poles hit the contour of integration are, other than
 $|m_1+m_2|<-\frac
 12$ and $|\overline m_1+\overline m_2|<-\frac 12$ for $Q^{w=0}$,
$min\left\{m_1+m_2,\overline m_1+\overline
 m_2\right\}<-\frac{k-1}{2}$
for $Q^{w=-1}$ and
$max\left\{m_1+m_2,\overline m_1+\overline m_2\right\}>\frac{k-1}{2}$
 for $Q^{w=+1}$.
For other
values of $j_1,j_2$ and $m_1,m_2$ the
OPE must be defined by analytic continuation. In the rest of this
section we perform this continuation.

To specifically display the contributions to (\ref{opec}) we have
to study the analytic structure of $Q^w$. We first consider the
simpler case
$w=\pm 1$ and we refer to the terms proportional to
$Q^{w=\pm 1}$ as {\it spectral flow non-preserving contributions} to the OPE. Then,
 we investigate $Q^{w=0}$ and obtain the {\it spectral flow preserving contributions}.

\subsection{Spectral flow non-preserving contributions}

Let us study the analytic structure of
$Q^{w=\pm 1}$ in (\ref{qw}).
The $m$-independent poles arising from the last
factor  are
 the same for both $w=\pm 1$ sectors
and
are explicitly given by
\begin{eqnarray}
     j_3 = \pm j^-_{21}+\frac k2 -1+p+q(k-2)\, ,\qquad  &&
j_3=\pm j^-_{21}-\frac k2 -p-q(k-2) \, ,\nonumber\\
        j_3=\pm j^+_{21}+\frac k2-1+p+q(k-2)\, ,\qquad &&
    j_3=\pm j^+_{21}-\frac k2-p-q(k-2)
\, ,\label{polos}
\end{eqnarray}
with $p,q = 0,1,2,\dots$
The $m$-dependent poles, instead, vary according to the spectral flow
sector.
However they
are connected through
 $(m,\overline m)\leftrightarrow(-m,-\overline m)$ and thus
going from
 $w= -1$ to  $w=+1$  involves the
change
${\cal D}^{-,\;w_i}_{j_i}
\otimes{\cal D}^{-,\;w_i}_{j_i}
\leftrightarrow
{\cal D}^{+,\;w_i}_{j_i}\otimes{\cal D}^{+,\;w_i}_{j_i}$.
Therefore we concentrate on the contributions from
$w=-1$.

 By abuse of
notation, from now on
we denote the states by the representations they belong to and we write only the
holomorphic sector for short, $e.g.$ when $\Phi_{m_i,\overline m_i}^{j_i,w_i}\in
{\cal D}^{+,w_i}_{j_i}\otimes{\cal D}^{+,w_i}_{j_i},
\, i=1,2$, we
 write the set of all possible operator products
$\Phi_{m_1,\overline m_1}^{j_1,w_1}\Phi_{m_2,\overline
m_2}^{j_2,w_2}$ for generic quantum numbers within these
representations as $ {\cal D}^{+,w_1}_{j_1}\times{\cal
D}^{+,w_2}_{j_2}$. 

Let us study the OPE of fields in all different
combinations of representations. First consider  the case
$\Phi_{m_i,\overline m_i}^{w_i,j_i}\in {\cal
C}_{j_i}^{\alpha_i,w_i} \otimes {\cal C}_{j_i}^{\alpha_i,w_i},
i=1,2$, $i.e.$
\newpage

\begin{itemize}
    \item ${\cal C}_{j_1}^{\alpha_1,\;w_1}\times\;
{\cal C}_{j_2}^{\alpha_2,\;w_2}$
\end{itemize}

The pole structure of $Q^{w=-1}$
 is represented in Figure 1.$a)$ for $min\left\{m_1+m_2,\,\overline m_1 +
\overline m_2\right\}<-\frac {k-1}{2}$. Recalling that
$m_3=m_1+m_2+\frac k2$, then
$min\left\{m_3,\overline m_3\right\}<\frac 12$, and therefore the poles
 from the factor $\frac{\Gamma(-j_3-\overline m_3)}{\Gamma(1+j_3+m_3)}$
are to the right of the  integration contour. Moreover, given that all
$m-$independent poles are to the right of the axis $\frac k2-1$
or to the left of $-\frac k2$, we conclude that
the OPE ${\cal
  C}_{j_1}^{\alpha_1,w_1}\times ~{\cal C}_{j_2}^{\alpha_2,w_2}$ receives no
 spectral flow
violating contributions from discrete representations when $min\{m_1+m_2,
\overline m_1+\overline m_2\}<-\frac{k-1}2$.

\bigskip

\medskip

\centerline{\psfig{figure=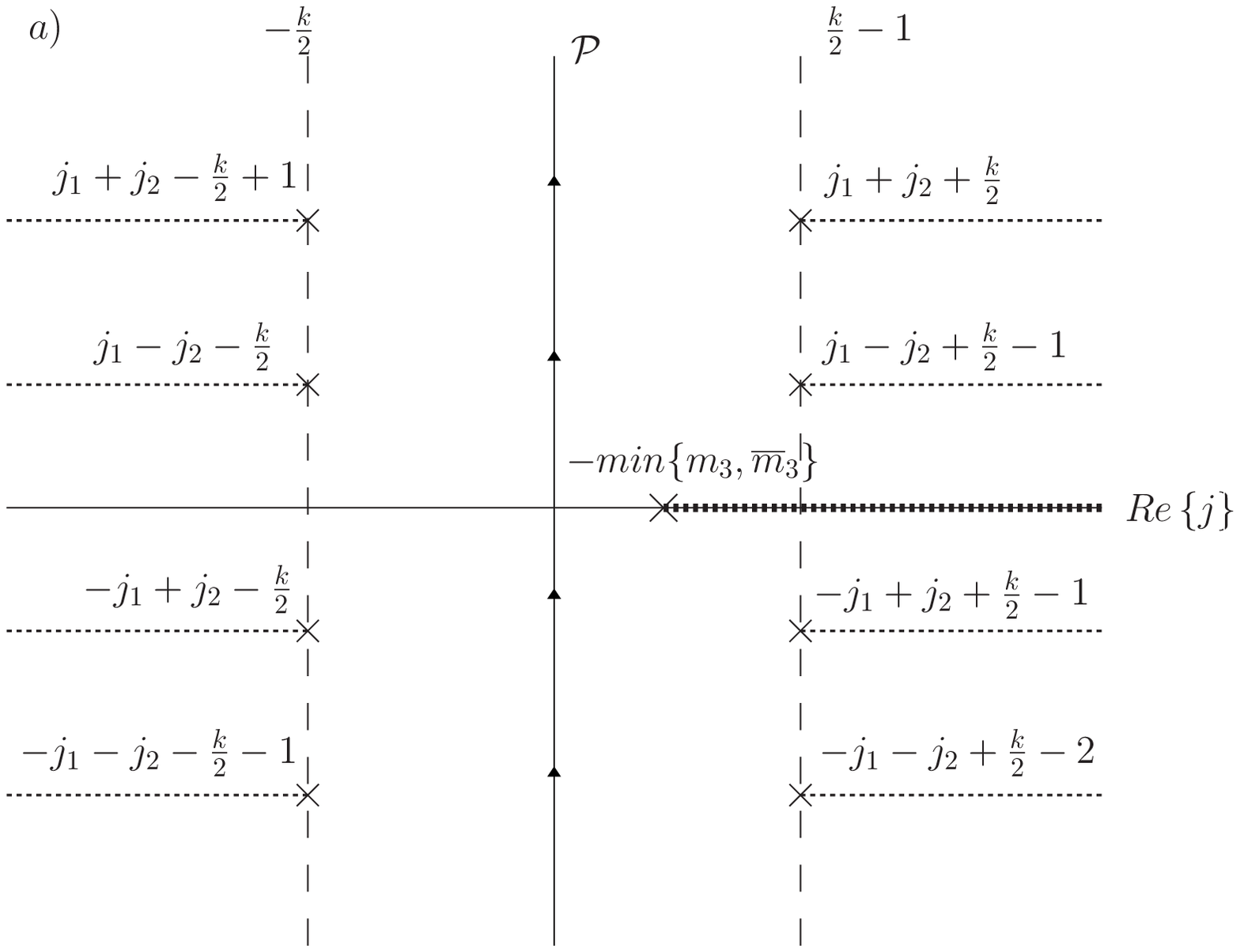,width=6.5cm}~~~~~~~~~~~~
\psfig{figure=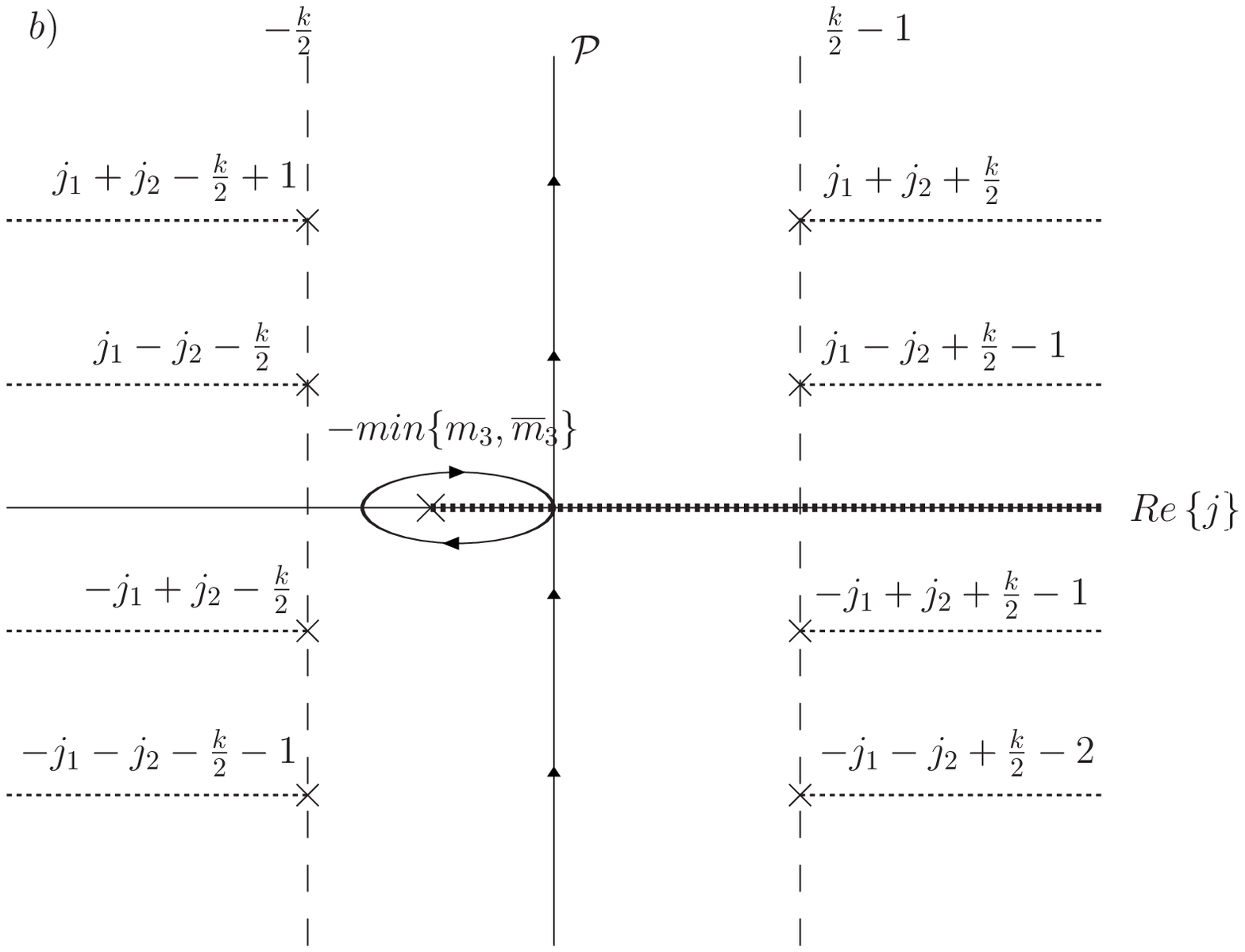,width=6.5cm}}
 {\footnotesize
{ Figure 1: Case ${\cal C}^{\alpha_1,w_1}_{j_1}\times{\cal
    C}^{\alpha_2,w_2}_{j_2}$.
The solid line indicates the integration
     contour ${\cal P}=-\frac 12+i{\mathbb R}$
in the $j_3$ complex plane. The dots above or below the real axis
     represent the ($j_1,j_2$)-dependent poles and those on the real axis
      correspond to  the $m-$dependent poles. The crosses are
the positions of the first poles in the series. $a)$
When
$m_1+m_2<-\frac{k-1}2$ or $\overline m_1+\overline
      m_2<-\frac{k-1}{2}$,
there are
no poles crossing the contour of integration. $b)$ When
$m_1+m_2>-\frac{k-1}2$ and $\overline m_1+\overline
      m_2>-\frac{k-1}{2}$,
 poles from the factor $\frac{\Gamma(-j_3-\overline m_3)}   {\Gamma(1+j_3+m_3)}$
cross the contour, indicating the contribution to
the OPE from states in discrete
     representations.}}
\bigskip

\medskip
Some poles cross the integration
contour when
$min\left\{m_1+m_2,\,\overline m_1 + \overline m_2\right\}>-\frac
{k-1}{2}$.
They are sketched in Figure $1.b)$ and  indicate contributions
from the
discrete series ${\cal D}_{j_3}^{+,w_3=w_1+w_2-1}$ with
$j_3=-min\left\{m_3,\overline m_3\right\}+n$,  $n=0,1,2,\dots$, and
such that $j_3<-\frac 12$. Since $Q^{w=\pm 1}$ does not vanish for
$j_3=-\frac 12 +i\mathbb R$ and $m_3$ not correlated with $j_3$,
 there are terms from ${\cal
  C}_{j_3}^{\alpha_3,w_3=w_1+w_2-1}$ in this OPE as well. Therefore we get

\bea
\left.{\cal C}^{\alpha_1,\;w_1}_{j_1}\times
{\cal C}^{\alpha_2,\;w_2}_{j_2}\right|_{|w|=1}
&=&\sum_{j_3<-\frac
  12} {\cal
    D}^{+,\;w_3=w_1+w_2-1}_{j_3}+\sum_{j_3<-\frac
  12} {\cal
    D}^{-,\;w_3=w_1+w_2+1}_{j_3}\nonumber\\
&&+\;\sum_{w=-1,1}
\int_{\mathcal{P}} dj_3\;{\cal
  C}^{\alpha_3,\;w_3=w_1+w_2+w}_{j_3}+\cdots ,\label{cc-1}
\eea
where  $|_{|w|= 1}$ denotes  that only spectral flow
non-preserving contributions are displayed in the right-hand side.
\begin{itemize}
    \item ${\cal C}_{j_1}^{\alpha_1,\;w_1}\times\;
{\cal D}_{j_2}^{\pm,\;w_2}$
\end{itemize}

To analyze this case, we need to perform the analytic continuation
for $j_2$ away from $-\frac 12+is_2$. When $is_2$ is continued to
the real interval $(-\frac{k-2}{2},\;0)$, the series of
$m-$independent poles changes as  shown in Figure
2. It is easy to see that these poles do not cross the contour of
integration. For instance, ${\rm
Re}\left\{j_1+j_2+\frac k2\right\}>0$, ${\rm
Re}\left\{j_1-j_2+\frac k2-1\right\}>\frac k2-1$, etc. Similarly as
in the previous case, only poles  from
$\frac{\Gamma(-j_3-\overline m_3)}{\Gamma(1+j_3+m_3)}$ can cross
the contour, but due to the factor
$\frac{\Gamma(1+j_2+\overline m_2)}{\Gamma(-j_2- m_2)}$ there are
contributions from the discrete series just for
$\Phi^{j_2,w_2}_{m_2,\overline m_2}\in {\cal D}^{-,w_2}_{j_2}\otimes
{\cal D}^{-,w_2}_{j_2}$. Therefore we get
\bea
\left.{\cal C}^{\alpha_1,\;w_1}_{j_1}
\times\;{\cal D}^{\pm,\;w_2}_{j_2}
\right|_{|w|=1}= \displaystyle  \int_{\mathcal{P}}
dj_3\;{\cal C}^{\alpha_3,\;w_3=w_1+w_2\pm1}_{j_3} + ~
\sum_{j_3<-\frac 12} {\cal D}^{\mp,\;w_3=w_1+w_2\pm1}_{j_3}+\cdots\,
.\label{d-c-1}
\eea

\centerline{\psfig{figure=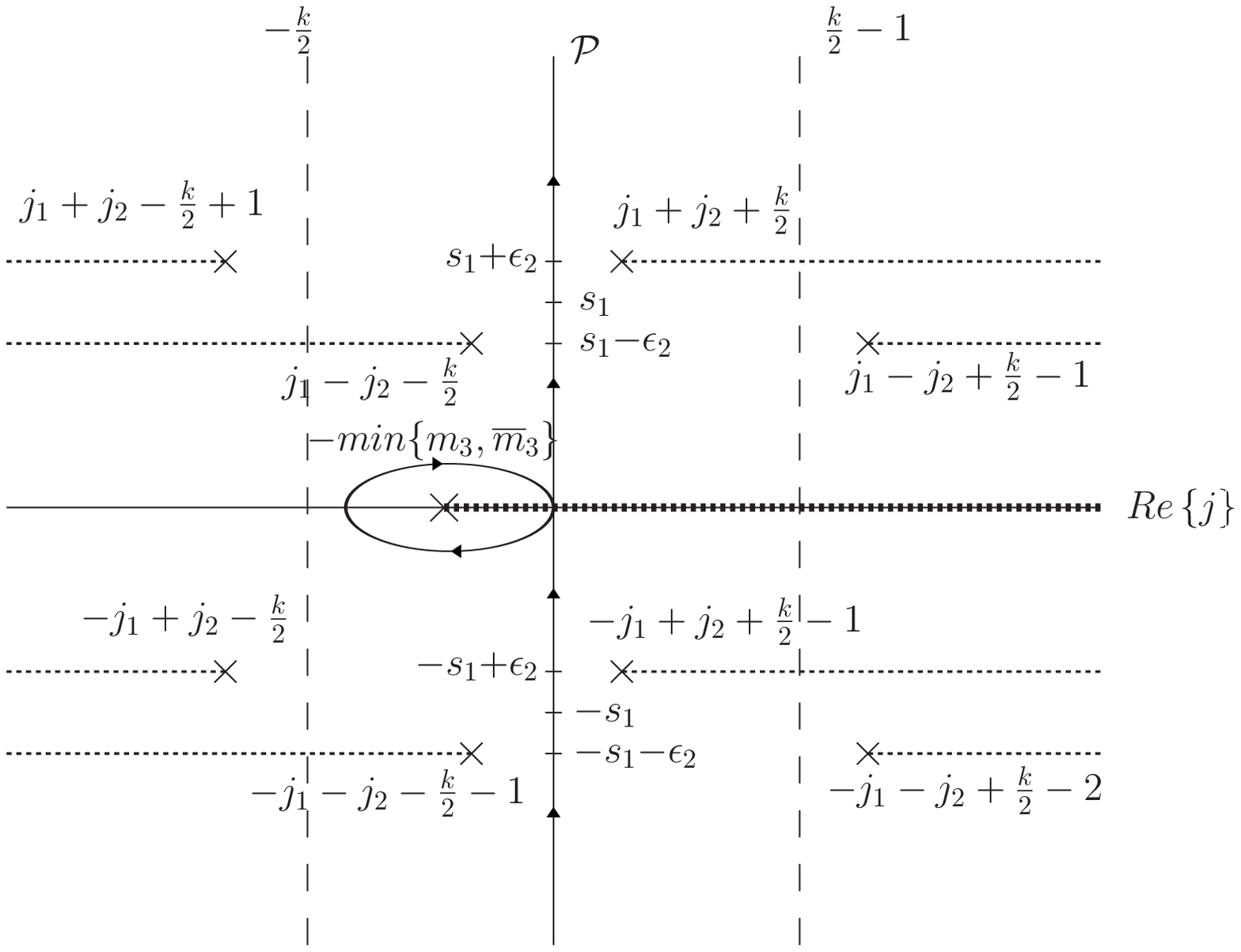,width=7cm}}
{\footnotesize Figure 2: Case ${\cal
      C}^{\alpha_1,w_1}_{j_1}\times{\cal D}^{\pm,w_2}_{j_2}$. Only $m-$dependent
      poles can cross
the contour of integration. This occurs when
 both $m_1+m_2$ and $\overline m_1+\overline m_2$ are
      larger
than $-\frac{k-1}{2}$. We have given $j_2$  an
infinitesimal imaginary part, $\epsilon_2$, to better display the
($j_1,j_2$)-dependent series of poles.}

\begin{itemize}
    \item ${\cal D}_{j_1}^{\pm,\;w_1}\times\;
{\cal D}_{j_2}^{\pm,\;w_2}$ and ${\cal D}_{j_1}^{\pm,\;w_1}\times\;
{\cal D}_{j_2}^{\mp,\;w_2}$

\end{itemize}

Let us first analytically continue both $j_1$ and $j_2$
to the interval $(-\frac{k-1}{2},\;-\frac 12)$, which is shown in Figure
3. The correct way to do this is to consider that both
$j_1$ and $j_2$
have an infinitesimal imaginary part, $\epsilon_1$ and $\epsilon_2$
respectively, which is sent  to zero after computing the integral.

The
$m-$independent poles cross the contour of integration only when
$j_1+j_2<-\frac{k+1}{2}$. However, due to the factors
$\frac{\Gamma(1+j_1+m_1)}{\Gamma(-j_1-\overline
m_1)}\frac{\Gamma(1+j_2+\overline m_2)}{\Gamma(-j_2- m_2)}$ 
in $Q^{w=-1}$, the contributions from these poles only survive  when
the quantum numbers of both
$\Phi^{j_1,w_1}_{m_1,\overline m_1}$ and $\Phi^{j_2,w_2}_{m_2,\overline
m_2}$ are in ${\cal D}_{j_i}^{-,w_i}\otimes{\cal D}_{j_i}^{-,w_i}, i=1,2$.
In this case,
the
poles at $j_3=j_1+j_2+\frac k2+n$ give  contributions
from
${\cal D}^{-,w_3=w_1+w_2-1}_{j_3}$. This may be seen
noticing that $j_3=m_1+m_2+\frac k2 +n_3=\overline m_1+\overline
m_2+\frac k2 +
\overline n_3$, with $n_3=n+n_1+n_2$ and $\overline n_3=n+\overline
n_1+\overline n_2$,
or using $m_3=m_1+m_2+\frac k2,\,\overline m_3=\overline m_1+\overline
m_2+\frac k2,$
so that $j_3=m_3+n_3=\overline m_3+\overline n_3$.
Instead, the contributions from the poles at $j_3=-j_1-j_2-\frac k2-1-n$
 seem to cancel due to the factor
$\frac{\Gamma(-j_3-\overline m_3)}{\Gamma(1+j_3+m_3)}$. However,
these zeros are cancelled because the operator  diverges. In
fact, using (\ref{rr}) and relabeling $j_3\rightarrow-1-j_3$, it is
straightforward to recover exactly the same contribution from the
poles at $j_3=j_1+j_2+\frac k2+n$. Obviously, this was expected
 as a consequence of the symmetry $j_3\leftrightarrow-1-j_3$
of the integrand in (\ref{opec}).

\bigskip

\centerline{
\psfig{figure=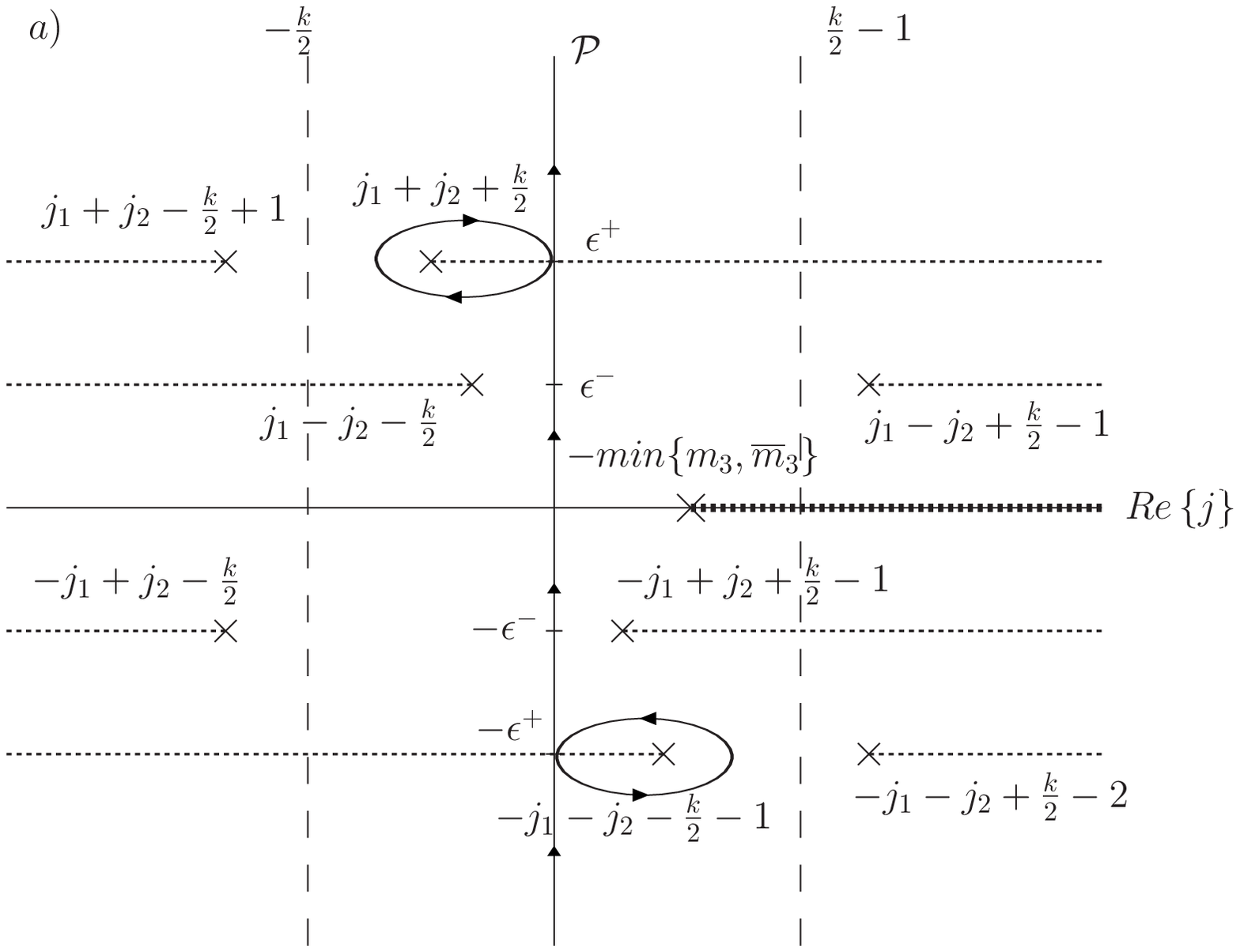,width=6.5cm}~~~~~~~~~~~~\psfig{figure=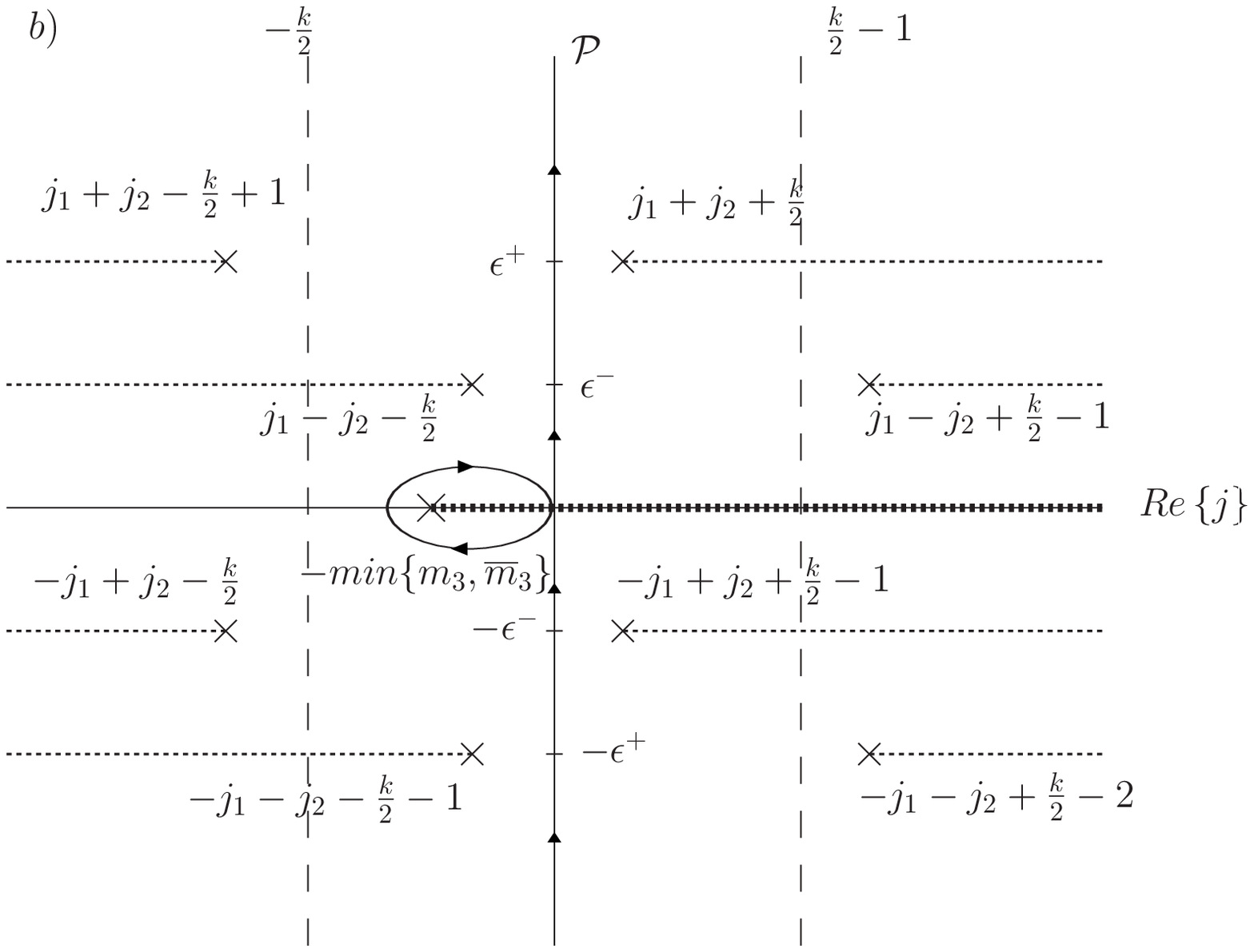,width=6.5cm}}
{\footnotesize Figure 3: Case
${\cal D}^{w_1}_{j_1}\times{\cal
       D}^{w_2}_{j_2}$.
Both $m-$dependent and
    $m-$independent poles can cross the contour of  integration. There
    are two
possibilities: 1)  ${\cal D}^{-,w_1}_{j_1}\times{\cal
       D}^{-,w_2}_{j_2}$.
When $j_1+j_2<-\frac{k+1}{2}$, only
$m-$independent poles can cross the contour, as shown in
Figure 3.$a$)
and when
$j_1+j_2>-\frac{k-1}{2}$, only $m-$dependent poles can  cross as shown in
Figure 3.$b$). 2) ${\cal D}^{\mp,w_1}_{j_1}\times{\cal
       D}^{\pm,w_2}_{j_2}$. 
Both
$m-$dependent and
$m-$independent poles can cross the contour  but only the former
survive after
taking the limit $\epsilon^+,\epsilon^-\rightarrow0$, where
$\epsilon^{\pm}=\epsilon_1\pm\epsilon_2$.}

\bigskip

Finally, the $m-$dependent poles give contributions from
${\cal D}^{+,w_3=w_1+w_2-1}_{j_3}$. Actually, when
$min\left\{m_3,\overline m_3\right\}>
\frac 12$
some of the $m-$dependent poles cross the  contour. Using $m-$conservation
it is not difficult to check that these contributions
fall inside the range (\ref{unitary}).

Let us continue the analysis, considering the OPE ${\cal D}^{\mp,w_1}_{j_1}\times{\cal
       D}^{\pm,w_2}_{j_2}$.
For instance, take the limiting case
$j_1=m_1+n_1+i\epsilon_1$ and $j_2=-m_2+n_2+i\epsilon_2$ with
$\epsilon_1,\;\epsilon_2\rightarrow0$. The factor
$\frac{\Gamma(1+j_2+\overline m_2)}{\Gamma(-j_2-m_2)}$ 
vanishes as a simple zero. However, some poles from the
series $j_3=j_2-j_1-\frac k2-n$ will overlap with
the $m-$dependent
poles. But because the $m-$independent simple poles are outside the
contour of integration, in the limit  $\epsilon_i\rightarrow0$ they may
cancel the simple zeros.
The way to compute this limit is determined by
the definition of
the three-point function. We
assume that a finite and nonzero term remains in the
limit \footnote{
In the limit $\epsilon_1,\epsilon_2\rightarrow0$,
$Res(Q^{w=-1})\sim\frac{\epsilon_2}{\epsilon_2-\epsilon_1}$.
The same ambiguity appears in the
three-point function including $\Phi^{j_1,w_1}_{m_1,\overline m_1}\in\mathcal
  D^{-,w_1}_{j_1}\otimes \mathcal D^{-,w_1}_{j_1}$,
$\Phi^{j_2,w_2}_{m_2,\overline m_2}\in \mathcal D^{+,w_2}_{j_2}\otimes
\mathcal D^{+,w_2}_{j_2}$, with
  $n_1\leq n_2$
such that $j_3=j_1-j_2-\frac k2-{\mathbb Z}_{n\ge 0}$.
The  resolution of this ambiguity requires an interpretation of the
divergences. 
The $w-$selection rules allow to
assume that a finite term survives  in the
limit. For instance, consider a generic three-point function 
$\langle{\cal D}_{j_1}^{-,w_1}{\cal D}^{+,w_2}_{j_2}
{\cal D}^{+,w_3}_{j_3}\rangle$ with $w_1+w_2+w_3=-1$.
According to (\ref{dd}) this is non-vanishing (for certain  values of
$j_i$, not determined from the $w-$selection rules). 
Indeed, the divergence from the $\delta^2(\sum_im_i-\frac k2)$ in
(\ref{3pointwind}) cancels the zero
from $\Gamma(-j_3-m_3)$ and then the pole in
$\widetilde C(1+j_i)\sim\frac{1}{\epsilon_2-\epsilon_1}$ must cancel
the zero from $\frac{\Gamma(1+j_2+\overline m_2)}{\Gamma(-j_2-\overline m_2)}
\sim\epsilon_2$, leaving a finite and non vanishing contribution.}.

Including the contributions
from continuous representations, we get the following  results:

\bea
\left.{\cal D}^{\pm,\;w_1}_{j_1}
\times ~
{\cal D}^{\pm,\;w_2}_{j_2}\right |_{|w|= 1}
&=& \int_{\mathcal{P}^+} dj_3 ~ {\cal C}^{\alpha_3,\;
w_3=w_1+w_2\pm1}_{j_3}
~ + ~\sum_{-j_1-j_2-\frac k2\le j_3< -\frac 12}
 {\cal D}^{\mp,\;w_3=w_1+w_2\pm 1}_{j_3}\nonumber\\
 && + ~
\sum_{j_1+j_2+\frac k2\le j_3< -\frac 12}
{\cal D}^{\pm,\;w_3=w_1+w_2\pm1}_{j_3}~ +\cdots .\label{lu}
\eea

\bea
\left .{\cal D}^{+,\;w_1}_{j_1}\times\; {\cal D}^{-,\;w_2}_{j_2}\right |_{|w|=1} &=&
\sum_{j_3<j_2-j_1-\frac k2}{\cal D}^{-,w_3=w_1+w_2+1}_{j_3}~ +
\sum_{j_3<j_1-j_2-\frac k2}{\cal D}^{+,w_3=w_1+w_2-1}_{j_3}~
+\cdots\nonumber\\
\label{d+td-}
\eea

\subsection{Spectral flow preserving contributions}

The analytic structure of  $Q^{w=0}(j_i;m_i,\overline
m_i)$ in (\ref{q0}) was studied in \cite{satoh}. Here we present the
analysis mainly to discuss some subtleties which are crucial to
perform the analytic continuation of $m_i, \overline m_i,
i=1,2$. Although our treatment of the $m-$dependent poles differs from
that followed in \cite{satoh}, we show in this section that the  results coincide.

The function $C(1+j_i)$ has zeros at $j_i=\frac{j-1}{2}$, $i=1,2,3$
and poles at $j=-j_1-j_2-j_3-2$,  $-1-j_1-j_2+j_3, ~ -1-j_1-j_3+j_2$, or $
-1-j_2-j_3+j_1$ where
$j:=p+q(k-2), -(p+1)-(q+1)(k-2)$, $p, q = 0, 1, 2, \cdots$.
To explore the behavior of the function $W$, we  use the
expression
\cite{satoh}
\begin{equation}
W\left[\begin{matrix}
j_1\,,\,j_2\,,\,j_3\cr  m_1, m_2,
m_3\cr\end{matrix}\right]=(i/2)^2\left [C^{12}
\overline P^{12}+C^{21}\overline P^{21}
\right ]\, ,\label{w}
\end{equation}
with $~~~~(i/2)^2P^{12}=s(j_1+m_1)s(j_2+m_2)C^{31}-s(j_2+m_2)s(m_1-j_2+j_3)C^{13}\,,$
\begin{eqnarray}
C^{12} &=&\frac{\Gamma(-N)\Gamma(1+j_3-m_3)}{\Gamma(-j_3-m_3)}
G\left[\begin{matrix}
    -m_3-j_3,\;-j_{13},\;1+m_2+j_2
\cr
                       -m_3-j_1+j_2+1,\;m_2-j_1-j_3\cr\end{matrix}\right]\, ,\nonumber\\
C^{31}&=&\frac{\Gamma(1+j_3+m_3)\Gamma(1+j_3-m_3)}{\Gamma(1+N)}
G\left[\begin{matrix}
    1+N,\;1+j_{1}+m_1,\;1-m_2+j_2
\cr
                       j_3+j_2+m_1+2,\;j_1+j_3-m_2+2\cr\end{matrix}\right]\, ,\nonumber\\
G\left[\begin{matrix} a,b,c\cr
    e,f\cr\end{matrix}\right]&=&\frac{\Gamma(a)\Gamma(b)
\Gamma(c)}
{\Gamma(e)\Gamma(f)}
F\left[\begin{matrix} a,b,c\cr e,f\cr\end{matrix}\right]
=\sum_{n=0}^{\infty}\frac 1{n!}\frac{\Gamma(a+n)\Gamma(b+n)
\Gamma(c+n)}
{\Gamma(e+n)\Gamma(f+n)\Gamma(n+1)}\,\label{Wfull} ,
\eea
and $N=1+j_1+j_2+j_3$, $s(x)=\sin(\pi x)$.
$\overline P^{ab}\,(\overline C^{ab})$ is obtained from $P^{ab}\,(C^{ab})$ by
 replacing $(m_i\rightarrow \overline m_i)$ and $P^{ba}\,(C^{ba})$ from
   $P^{ab}\, (C^{ab})$ by  changing $(j_1,m_1
\leftrightarrow j_2,m_2)$ and
$F\left[\begin{matrix} a,b,c\cr e,f\cr\end{matrix}\right]={}_3F_2(a,b,c;e,f;1)$.
An equivalent expression for $W$ which will be
useful below is the following \cite{satoh}
\bea
W\left[\begin{matrix}
j_1\,,\,j_2\,,\,j_3\cr m_1, m_2, m_3\cr\end{matrix}\right]=D_1C^{12}\overline C^{12}+
D_2C^{21}\overline C^{21}+D_3[C^{12}\overline C^{21}+C^{21}\overline
  C^{12}]\, ,
\label{ww}
\eea
where
\bea
D_1 &=& \frac{s(j_2+m_2)s(j_{13})}{s(j_1-m_1)s(j_2-m_2)s(j_3+m_3)}
[s(j_1+m_1)s(j_1-m_1)s(j_2+m_2)\nonumber\\
&&\; -
s(j_2-m_2)s(j_2-j_3-m_1)s(j_2+j_3-m_1)]\, ,\nonumber\\
D_2&=&D_1(j_1,m_1\leftrightarrow j_2,m_2)\, ,\nonumber\\
D_3&=&-\frac{s(j_{13})s(j_{23})s(j_1+m_1)s(j_2+m_2)s(j_1+j_2+m_3)}
{s(j_1-m_1)s(j_2-m_2)s(j_3+m_3)}\,\label{D} .
\eea

Studying the analytic structure of $Q^{w=0}$ is a difficult task as a
consequence of the complicated form
of  $W$.
The analysis greatly simplifies  when analytically continuing the
quantum numbers of one  operator to those of a
discrete representation. Indeed,
when
$j_1=-m_1+n_1=-\overline m_1+\overline n_1,$ and
$n_1,\overline n_1= 0,1,2,\cdots$,
 $W{\scriptsize\left[\begin{matrix}
j_1\,,\,j_2\,,\,j_3\cr  m_1, m_2, m_3\cr\end{matrix}\right]}$ reduces to
$W_1=D_1C^{12}\overline C^{12}$
\cite{satoh}, $i.e.$
\begin{equation}
W_1\left[\begin{matrix}
j_1\,,\,j_2\,,\,j_3\cr  m_1, m_2, m_3\cr\end{matrix}\right]=
\frac{(-)^{m_3-\overline m_3+\overline n_1}\pi^2 \gamma(-N)
}{\gamma(-2j_1)\gamma(1+j_{12})\gamma(1+j_{13})}\frac{\Gamma(1+j_3-m_3)
\Gamma(1+j_3-\overline
m_3)}{\Gamma(1+j_3-m_3-n_1)\Gamma(1+j_3-\overline
 m_3-\overline n_1)}\nonumber
\end{equation}
\begin{equation}
~~~~~~~~~~~~~
\times \prod_{i=2,3}\frac{\Gamma(1+j_i+m_i)}{\Gamma(-j_i-\overline m_i)}F\left
[\begin{matrix} -n_1,-j_{12},1+j_{23}\cr
-2j_1,1+j_3-m_3-n_1\cr\end{matrix}\right]
F\left[\begin{matrix} -\overline n_1,-j_{12},1+j_{23}\cr
-2j_1,1+j_3-\overline m_3-\overline
n_1\cr\end{matrix}\right]\label{w1}\, .
\end{equation}

It is easy to see that
\bea
& &\frac{\Gamma(1+j_3-m_3)}{\Gamma(1+j_3-m_3-n_1)}F\left[\begin{matrix}
-n_1,-j_{12},1+j_{23}\cr-2j_1,1+j_3-m_3-n_1\cr\end{matrix}\right]=\cr\cr
& &\sum_{n=0}^{n_1}\frac{(-)^n n_1!}{n!(n_1-n)!}\frac{\Gamma(n-j_{12})}
{\Gamma(-j_{12})} \frac{\Gamma(n+1+j_{23})}{\Gamma(1+j_{23})} \frac{\Gamma(-2j_1)}
{\Gamma(n-2j_1)}\frac{\Gamma(1+j_3-m_3)}{\Gamma(n+1+j_3-m_3-n_1)}\,
 .
\label{Fn1} \eea
Recall that the OPE
involves the function $W{\scriptsize\left[\begin{matrix}
j_1\,,\,j_2\,,\,j_3\cr  m_1, m_2, -m_3\cr\end{matrix}\right]}$ and
then the change $(m_3,\overline m_3)\rightarrow (-m_3,-\overline
m_3)$ is required in the above expressions to analyze $Q^{w=0}$.
 Thus, for generic $2j_i\notin\mathbb{Z}$, the
poles and zeros of $Q^{w=0}(j_i;m_i,\overline m_i)$ are contained
in

\begin{equation}
C(1+j_i) \frac{\gamma(-1-j_1-j_2-j_3)}
{\gamma(1+j_{12})\gamma(1+j_{13})}
\frac{\Gamma(1+m_2+j_2)\Gamma(-m_3-j_3)} {\Gamma(-\overline
m_2-j_2)\Gamma(1+\overline m_3+j_3)}\, , \label{q0a}
\end{equation}
plus possible additional zeros in (\ref{Fn1}) and its antiholomorphic
equivalent expression (see appendix A.1).
The $(j_1,j_2)-$dependent poles in (\ref{q0a}) are at
$j_3=\pm j_{21}^{\pm}+p+(q+1)(k-2)$, $\pm j_{21}^{\pm}-(p+1)-q(k-2)$,
$\mp j_{21}^{\pm}+p+q(k-2)$, $\mp j_{21}^{\pm}-(p+1)-(q+1)(k-2)$. There are also
 zeros at $1+2j_i=p+q(k-2),\;-(p+1)-(q+1)(k-2)$, $i=1,2,3$.

Let us first
consider  $\Phi_{m_1,\overline m_1}^{w_1,j_1}\in
{\cal D}_{j_1}^{+,w_1}
\otimes {\cal D}_{j_1}^{+,w_1}$
 and note that when
 $\Phi_{m_1,\overline m_1}^{w_1,j_1}\in
{\cal D}_{j_1}^{-,w_1}
\otimes {\cal D}_{j_1}^{-,w_1}$
the OPE follows directly
 using the symmetry of the spectral flow conserving two- and three-point
functions under
$(m_i,\overline m_i)\leftrightarrow(-m_i,-\overline m_i), \forall ~
i=1,2,3$.\footnote {
~This symmetry follows directly from the integral expression for
 $W{\scriptsize\left[\begin{matrix}
j_1\,,\,j_2\,,\,j_3\cr  m_1, m_2, m_3\cr\end{matrix}\right]}$
performing the change of variables ($x_i,\overline
x_i)\rightarrow(x_i^{-1},\overline x_i^{-1})$ in (\ref{w1i}).}

\begin{itemize}
    \item ${\cal D}_{j_1}^{\pm,\;w_1}\times\;
{\cal C}_{j_2}^{\alpha_2,\;w_2}$
\end{itemize}

Consider
$j_1=-m_1+n_1+i\epsilon_1$ with $n_i\in{\mathbb Z}_{\ge 0}$ and
 $\epsilon_1$ an infinitesimal
positive number, and $j_2=-\frac 12 +is_2$
not correlated with $m_2$.
In this case,
$W{\scriptsize\left[\begin{matrix}
j_1\,,\,j_2\,,\,j_3\cr  m_1, m_2, m_3\cr\end{matrix}\right]}\approx
W_1{\scriptsize\left[\begin{matrix}
j_1\,,\,j_2\,,\,j_3\cr  m_1, m_2, m_3\cr\end{matrix}\right]}$.

The $m-$independent poles are to the right or to the left
of the contour of integration as  sketched in Figure 4.$a$).
If $min\left\{m_3,\overline
  m_3\right\}<\frac12$,
none of the $m-$dependent poles cross the contour, implying that only continuous
  series contribute to
the spectral flow conserving terms of the OPE
$ {\cal D}^{+,w_1}_{j_1}
\times
{\cal C}^{\alpha_2,w_2}_{j_2}$.   On the
  other hand if
$min\left\{m_3,\overline m_3\right\}>\frac12$, this OPE also receives
contributions
from ${\cal  D}^{+,w_3=w_1+w_2}_{j_3}$.
 Note that when $j_1\approx m_1+n_1$, $W{\scriptsize\left[\begin{matrix}
j_1\,,\,j_2\,,\,j_3\cr  m_1, m_2, m_3\cr\end{matrix}\right]}\approx
W_1{\scriptsize\left[\begin{matrix}
j_1\,,\,j_2\,,\,j_3\cr  -m_1, -m_2, -m_3\cr\end{matrix}\right]}$,
which implies that the spectral flow conserving terms in the OPE
${\cal D}^{-,w_1}_{j_1}\times
{\cal C}^{\alpha_2, w_2}_{j_2}$ contain  contributions from
the continuous representations as well as from
${\cal D}^{-,w_3}_{j_3}$ when $max\left\{m_3,\overline
m_3\right\}<-\frac12$. So we find
\bea
\left.{\cal D}^{\pm,\;w_1}_{j_1}\times\;
{\cal C}^{\alpha_2,\;w_2}_{j_2}\right|_{w=0}=\;\int_{\mathcal{P}}
dj_3\, {\cal C}_{j_3}^{\alpha_3,\;w_3=w_1+w_2}\;+\;
\sum_{j_3<-1/2}{\cal D}^{\pm,\;w_3=w_1+w_2}_{j_3}+\cdots \, .\label{dpmc0}
\eea

\begin{itemize}
    \item ${\cal D}_{j_1}^{\pm,\;w_1}\otimes\;
{\cal D}_{j_2}^{\mp,\;w_2}$ and ${\cal D}_{j_1}^{\mp,\;w_1}\otimes\;
{\cal D}_{j_2}^{\mp,\;w_2}$
\end{itemize}

When $j_2$ is
continued to $(-\frac{k-1}{2}+i\epsilon_2,-\frac12+i\epsilon_2)$,
$\epsilon_2$ being an infinitesimal positive number, $W$ is again
well approximated by $W_1$ as long as $j_2\neq-m_2+n_2+i\epsilon_2,
-\overline m_2+\overline n_2+i\epsilon_2$.
Otherwise, one also
has to consider $W_2\equiv D_{2}C^{21}\overline C^{21}$, but the result
coincides exactly with the one obtained using $W_1$, so we
restrict to this.
Two $m-$independent series of poles may cross the
contour of integration: $j_3=j_1-j_2-1-p-q(k-2)$ and
$j_3=j_2-j_1+p+q(k-2)$, both with $q=0$. The former has
$j_3>-\frac12$ and the latter,  $j_3<-\frac12$.
  The $m-$dependent poles in $Q^{w=0}$ arise from
$\frac{\Gamma(-j_3-\overline m_3)}{\Gamma(1+j_3+m_3)}$.
When $j_2=-m_2+n_2+i\epsilon_2$, because of the factor
$\Gamma(-j_2-m_2)^{-1}$, only $m-$dependent
poles give contributions from discrete series. To see this, consider
the $m-$independent
poles at
$j_3=j_1+j_2-p-q(k-2)$.
These  are outside the contour of integration and in the limit
$\epsilon_1,\epsilon_2\rightarrow0$ some of them may overlap with the
$m-$dependent ones.
Again,  one may argue that this limit leaves a finite and
non-vanishing factor.
\medskip

\centerline{
\psfig{figure=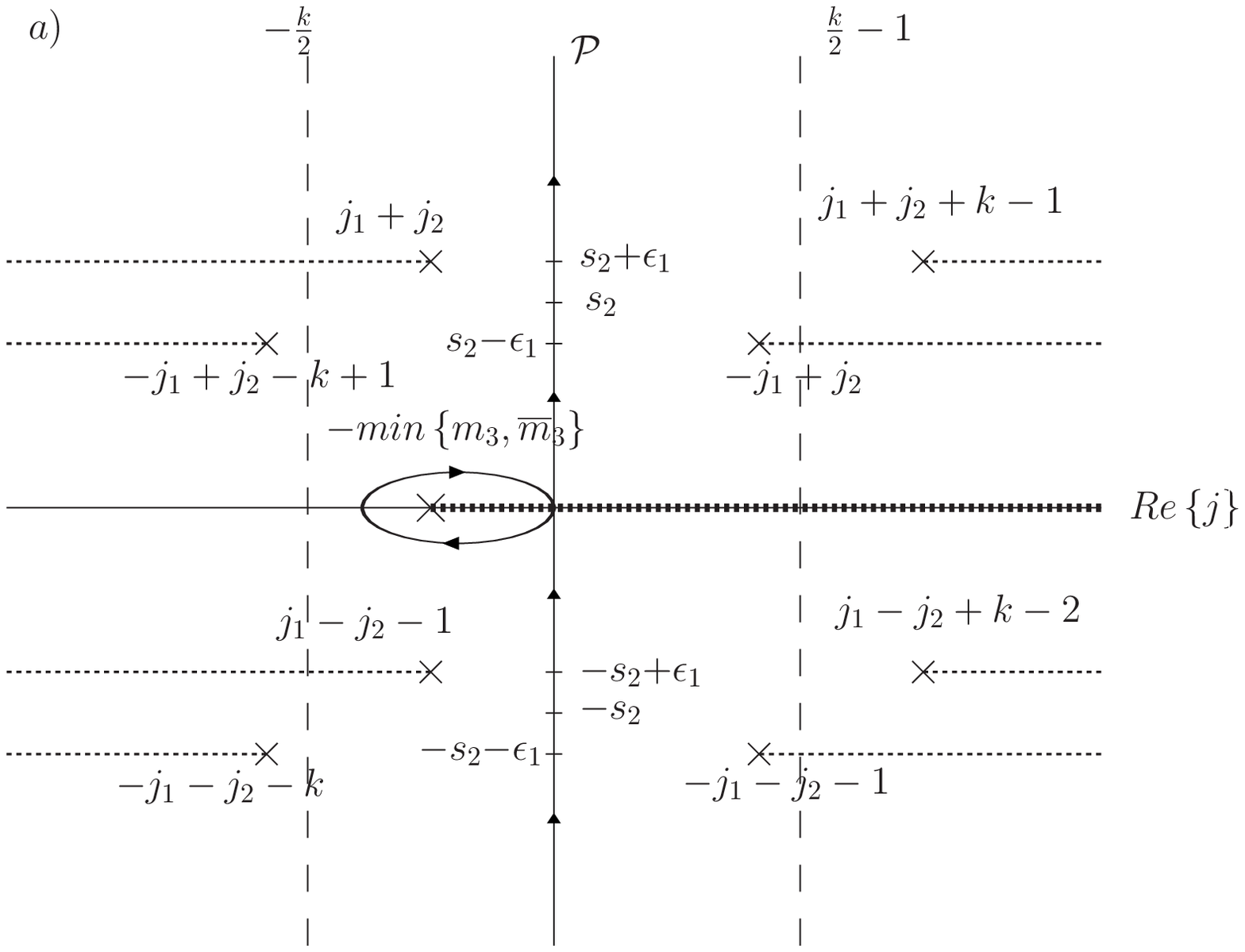,width=6.5cm}~~~~~~~~~~~~\psfig{figure=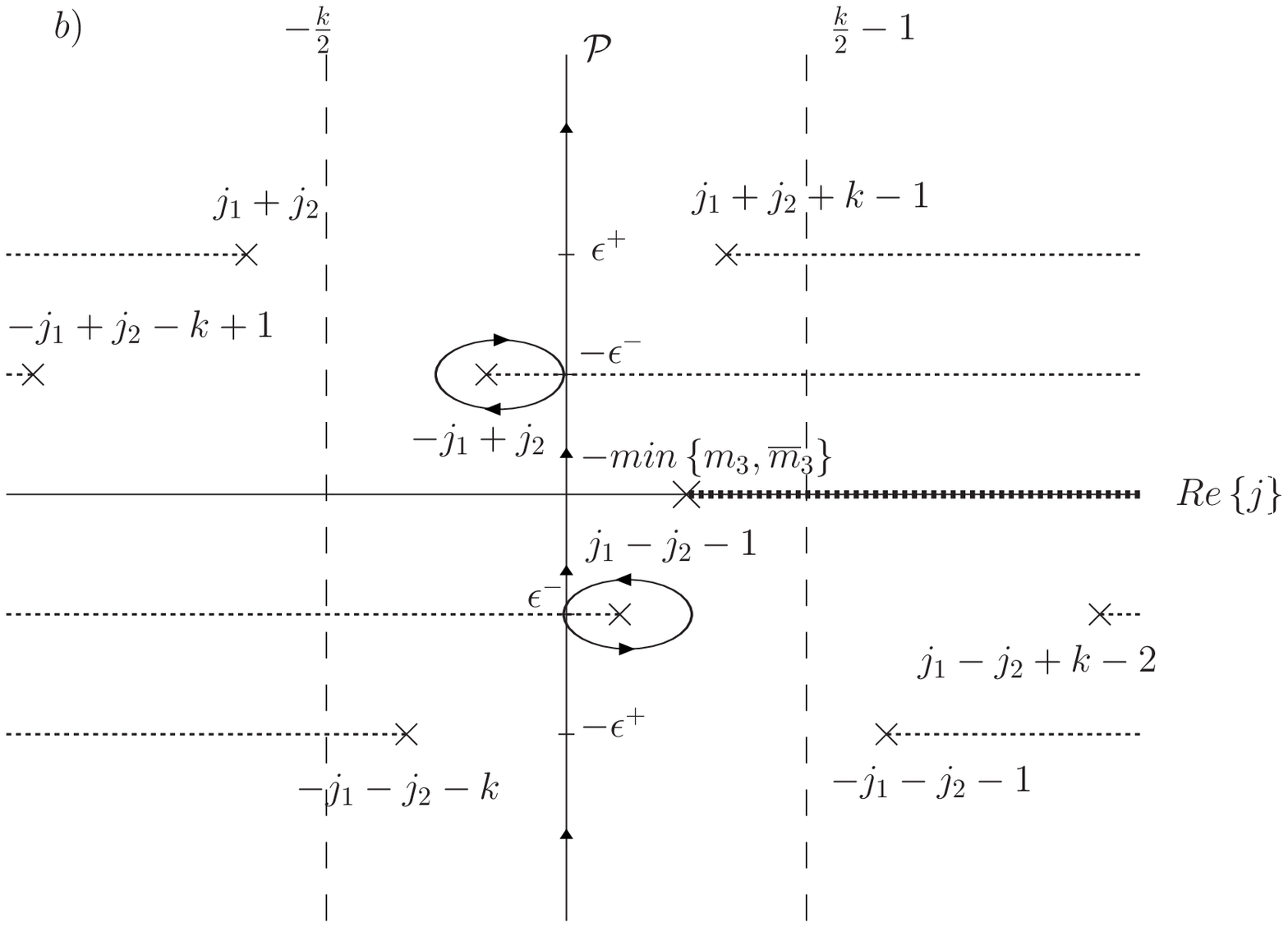,width=6.5cm}}

{\footnotesize Figure 4:
Analytic continuation of $Q^{w=0}$ for
($j_1,j_2$)-values away from the axis $-\frac 12+i\mathbb R$,
using $W_1$ instead of W.
 In  $4.a)$ $j_2=-\frac 12+is_2$
and only $m-$dependent poles can cross the contour of integration.
In $4.b)$ $-\frac{k-1}{2}<j_2<-\frac 12$ was considered. While
$m-$independent poles only cross the contour when $j_2<j_1$,
$m-$dependent poles can cross independently of the values of
$j_1,j_2$, but they are annihilated unless $j_2>j_1$. }
\bigskip

When $j_2=m_2+n_2+i\epsilon_2$, at first sight there are no
zeros.
 If $j_2-j_1<-\frac 12$, some poles with $q=0$ in the series
 $j_3=j_2-j_1+p+q(k-2)$ and
$j_3=j_1-j_2-1-p-q(k-2)$ cross the contour, as  shown in Figure
4.$b$). Using the relation between $j_i$ and $m_i$ and
$m-$conservation, it follows that the former poles
can be rewritten as $j_3=m_3+n_3=\overline m_3
+\overline n_3$, where $n_3=n_2-n_1+p$ and $\overline n_3=\overline
n_2-\overline n_1+p$. Obviously, if $n_2\geq n_1$ and $\overline
n_2\geq \overline n_1$ all the residues picked up by the contour
deformation
imply contributions to
the OPE from ${\cal D}^{-,w_3=w_1+w_2}_{j_3}$. When
$n_2<n_1$ or $\overline n_2<\overline n_1$,
only those values of $p$ for which both $n_3$ and $\overline n_3$
are non-negative integers remain after taking the limit
$\epsilon_1,\epsilon_2\rightarrow 0$. This is
because of extra zeros appearing in $W_1$ which are not explicit in
(\ref{w1})
(see appendix A.1).
 Using the results in the Appendix and the identity (\ref{rr}) it is
 straightforward to see that the latter  series of poles give the same contributions.

  The poles at
 $j_3=-min\left\{m_3,\overline m_3\right\}+n_3$ may cross the contour.
 If this happens they overlap with the $m-$independent poles. But
 there are
double zeros cancelling these contributions.

 If $j_2-j_1>-\frac 12$, only $m-$dependent poles may cross the
 contour. But they give
contributions only if they do not overlap with the poles at
 $j_3=j_1-j_2-1-n$, again because of the presence of double zeros. 
Therefore, these contributions remain only for
 $j_3\geq j_1-j_2$.

Putting all together we get
\bea
 \left.{\cal D}^{+,\; w_1}_{j_1}\times\; {\cal D}^{-,\;w_2}_{j_2}\right|_{w=0} &=&
\int_{\mathcal{P}} dj_3~
{\cal C}_{j_3}^{\alpha_3,\;w_3=w_1+w_2}\;
+\displaystyle\sum_{j_2-j_1\leq j_3<-\frac 12
}{\cal D}^{-,\;w_3=w_1+w_2}_{j_3}\; \nonumber\\
&&~~ + ~
\displaystyle\sum_{ j_1-j_2\leq j_3<-\frac
12}{\cal D}^{+,\;w_3=w_1+w_2}_{j_3}+\cdots \, ,\label{d+1d-20}\\
    \left.{\cal D}^{\pm,\;w_1}_{j_1}\times\;{\cal D}^{\pm,\;w_2}_{j_2}\right|_{w=0}&=&
    \displaystyle\sum_{j_3\leq j_1+j_2} {\cal
    D}^{\pm,\;w_3=w_1+w_2}_{j_3}+\cdots\, .
\label{ddis}
\eea

\begin{itemize}
    \item
${\cal C}_{j_1}^{\alpha_1,\;w_1}\times\;
{\cal C}_{j_2}^{\alpha_2,\;w_2}$
\end{itemize}

The zero and pole structure of $Q^{w=0}$  is given by
\bea
Q^{w=0}(j_i;m_i,\overline m_i) &\sim & C(1+j_i)
\frac{\gamma(-N)}{s(\overline m_3+j_3)}
G\left[\begin{matrix}
    m_3-j_3,\;-j_{13},\;1+m_2+j_2
\cr
                       m_3-j_1+j_2+1,\;m_2-j_1-j_3\cr\end{matrix}\right]\cr\cr
&&\times ~
\left\{s(\overline m_1+j_1)G\left[\begin{matrix} 1+N,\;1+\overline
    m_1+j_1,\;1-
\overline m_2+j_2\cr
                       2+\overline m_1+j_2+j_3,\;2-\overline
               m_2+j_1+j_3\cr\end{matrix}\right]
\right.\cr\cr&& ~~~ - ~ \left.
 s(\overline m_1-j_2+j_3) G\left[\begin{matrix} 1+N,\;1+\overline
 m_2+j_2,\;1-\overline m_1+j_1\cr
                       2+\overline m_2+j_1+j_3,\;2-\overline
               m_1+j_2+j_3\cr\end{matrix}\right]
\right\}\nonumber\\
&& + ~
(j_1,m_1,\overline m_1)\leftrightarrow(j_2,m_2,\overline m_2)\, .\nonumber \eea

$G{\scriptsize\left[\begin{matrix} a,b,c\cr
e,f\cr\end{matrix}\right]}$ has simple poles at
$a,b,c=0,-1,-2,\dots$ as well as at $u=e+f-a-b-c=0,-1,-2,\dots$, 
if $a,b, c\ne 0,-1,-2,\cdots$. So, a direct analysis leads to the conclusion that,
 when
$j_i=-\frac 12+is_i,i=1,2$, the
poles are contained in the following
expression: \bea
C(1+j_i)
\Gamma(-N)[\Gamma(-j_{12})]^2\Gamma(-j_{13})\Gamma(-j_{23})\Gamma(-j_3+m_3)\Gamma(-j_3-
\overline m_3)\Gamma(1+j_3+\overline m_3)\, .\label{wcxc1} \eea 
Instead, if one looks for poles in $Q^{w=0}$ using
(\ref{ww}), they seem to be those contained in \bea
C(1+j_i)
[\Gamma(-N)\Gamma(-j_{12})\Gamma(-j_{13})\Gamma(-j_{23})]^2\Gamma(-j_3+m_3)
\Gamma(-j_3-\overline m_3)\Gamma(1+j_3+\overline m_3)\,
.\label{wcxc2} \eea These different behaviors in the
$(j_1,j_2)-$dependent poles suggest that one must be very careful
when  analyzing the analytic structure of $Q^{w=0}$. The $(m_3,\overline m_3)-$dependent
poles  coincide in both expressions. However, the
symmetries of $W$ imply that for generic $j_1,j_2$ and $m_1,m_2$,
the $m_3-$dependent poles must be symmetric under
$m_3\leftrightarrow\overline m_3$ as well as under $(m_3,\overline
m_3)\leftrightarrow (-m_3,-\overline m_3)$, and this does not seem to be the
case in the expressions above.

  This puzzle is a consequence of the intricate functional form of $W$.
Extra zeros may be hidden. Actually,
let us
 show that the correct behavior of $Q^{w=0}$
  must be of the form
\footnote{~The pole structure of this expression
  is obviously symmetric
under $m_3\leftrightarrow\overline m_3$ as well as under
  $(m_3,\overline m_3)
\leftrightarrow (-m_3,-\overline m_3)$.}

\bea Q^{w=0}\sim \frac{\Gamma(-j_3-m_3)\Gamma(-j_3+\overline
m_3)}{\Gamma(1+j_3-m_3)\Gamma(1+j_3+\overline m_3)}\, ,\label{wcxc3}
\eea
for generic $j_1,j_2$ and for $m_1,m_2$ not correlated with them,
up to regular and non-vanishing contributions for
$j_3=\pm m_3+n_3=\pm\overline m_3+\overline n_3$, with
$n_3,\overline n_3\in \mathbb Z$.

To check (\ref{wcxc3}), let us consider
$j_3=-m_3+n_3+i\epsilon_3=-\overline m_3+\overline n_3+i
\epsilon_3$, with  $\epsilon_3$
an infinitesimal number. Using (\ref{ww}) with the relabeling
$1\leftrightarrow 3$, only a term like $D_1$  remains in $W$
 because the other terms behave as $\epsilon_3$
and there are no extra
divergences  to cancel the zeros when $\epsilon_3\rightarrow 0$.
Then, $W$ behaves as $W_1$ in (\ref{w1}), with
the relabeling
discussed above.
The factor ${\scriptsize\displaystyle \frac{\Gamma(1+j_1-m_1)}{\Gamma(1+j_1-m_1-n_3)}
F\left[\begin{matrix} -n_3,-j_{23},1+j_{12}\cr
-2j_3,1+j_1-m_1-n_3\cr\end{matrix}\right]}$ and the similar antiholomorphic
 one have no poles or zeros when $j_1$ and $m_1$ are not correlated.
  So, we conclude that for $j_3=-m_3+q_3+i\epsilon_3=-\overline
  m_3+\overline q_3
+i\epsilon_3$, $W$ has no $m_3-$dependent poles or zeros, and then
\bea
Q^{w=0}\sim
C(1+j_i)\gamma(-N)\gamma(-j_{23})\gamma(-j_{13})\frac{\Gamma(-j_3-m_3)
\Gamma(-j_3+\overline m_3)}{\Gamma(1+j_3-m_3)\Gamma(1+j_3+\overline
  m_3)}\, .
\eea

Using the symmetry  $(m_i,\overline m_i)\leftrightarrow
(-m_i,-\overline m_i)$ of $W$, it is straightforward to deduce that the same
behavior
is obtained for $j_3=m_3+n_3+i\epsilon_3=\overline m_3+\overline n_3+i\epsilon_3$.

We may now analyze the OPE ${\cal C}_{j_1}^{\alpha_1,\;w_1}\times\;
{\cal C}_{j_2}^{\alpha_2,\;w_2}$.
A sum over continuous
representations appears because  $Q^{w=0}$ does not vanish for
$j_3=-\frac 12+is_3$ when $s_3$ is a real number. On the other hand,
the expression (\ref{wcxc3}) shows that there are no contributions from
discrete representations
provided  $min\left\{m_1+m_2,\overline m_1+\overline
m_2\right\}<\frac 12$ and $max\left\{m_1+m_2,\overline m_1+\overline
m_2\right\}>-\frac 12$. Obviously both bounds cannot be violated at
the same time. When the first one is violated, operators belonging
to  spectral flow images of  lowest-weight representations
contribute to the OPE. On the contrary, when the second bound is not
satisfied, operators in spectral flow images of  highest-weight
representations appear in the OPE.

Extra poles could possibly appear in the $m-$basis implying contributions
from operators not belonging to ${\cal C}_j^{\alpha, w}$ or ${\cal
  D}_j^{\pm, w}$ 
representations.
However, the poles of ${}_3F_2$ are well known and  no other
than those in
(\ref{wcxc1}) and (\ref{wcxc2})   appear in $W$. Instead, there could
be extra
zeros cancelling
certain poles  as a consequence of particular
combinations of the arguments in ${}_3F_2$. As we have shown, these
possible  zeros cannot cancel the $m_3-$dependent poles.  This
information supports the conclusion that the OPE is closed among
${\cal C}_j^{\alpha, w}$ and ${\cal
  D}_j^{\pm, w}$ 
representations.

Finally, we want to remark the importance of a relation
like (\ref{wcxc3}), because the other expressions (\ref{wcxc1}) and
(\ref{wcxc2}) do not admit a definition of the OPE as  analytic
continuation since the $m_3-$dependent poles do not seem to begin (or
end) at a given
point.

Therefore, we conclude that the $w-$conserving contributions to the
OPE of two
continuous representations are the following:
\bea
\left.{\cal C}_{j_1}^{\alpha_1,\;w_1}\times\;
{\cal C}_{j_2}^{\alpha_2,\;w_2}\right|_{w=0} \sim
\int_{\mathcal{P}} dj_3\, {\cal C}_{j_3}^{\alpha_3,\;w_3=w_1+w_2}
&+& \sum_{j_3<-\frac12} {\cal D}^{+,\;w_3=w_1+w_2}_{j_3}
+\;\sum_{j_3<-\frac 12}
{\cal D}^{-,\;w_3=w_1+w_2}_{j_3}\, ,\nonumber\\\label{cc0}
\eea
up to descendants.
Note that, in a particular OPE with  $m_i,\overline m_i$ fixed, only
one of the discrete series contributes, depending on the signs of
 $m_3,\overline m_3$.

Collecting all the results,  
 the OPE for primary
fields and their
spectral flow images in the spectrum of the $SL(2,\mathbb R)$ WZNW
 model are the following:

\bea
{\cal D}^{\pm,\;w_1}_{j_1}
\times ~
{\cal D}^{\pm,\;w_2}_{j_2}
&=&
\sum_{ j_3\le j_1+j_2}
 {\cal D}^{\pm,\;w_3=w_1+w_2}_{j_3}+\sum_{-j_1-j_2-\frac k2\le j_3< -\frac 12}
 {\cal D}^{\mp,\;w_3=w_1+w_2\pm 1}_{j_3}
\nonumber\\
& +&
\sum_{j_1+j_2+\frac k2\le j_3< -\frac 12 }
  {\cal D}^{\pm,\;w_3=w_1+w_2\pm 1}_{j_3}+
\int_{\mathcal{P}} dj_3 ~ {\cal C}^{\alpha_3,\;
w_3=w_1+w_2\pm 1}_{j_3}+\cdots . \label{1}\\
 {\cal D}^{+,\; w_1}_{j_1}\times\; {\cal D}^{-,\;w_2}_{j_2} &=&
\displaystyle\sum_{ j_1-j_2\leq j_3<-\frac
12}{\cal D}^{+,\;w_3=w_1+w_2}_{j_3}
+\displaystyle\sum_{j_2-j_1\leq j_3<-\frac 12
}{\cal D}^{-,\;w_3=w_1+w_2}_{j_3}+
\nonumber\\
&&+\displaystyle\sum_{ j_3\le j_2-j_1-\frac k2}
{\cal D}^{-,\;w_3=w_1+w_2+1}_{j_3}+\displaystyle\sum_{j_3\le j_1-j_2-\frac k2
}{\cal D}^{+,\;w_3=w_1+w_2-1}_{j_3}\nonumber\\
&&+ ~\int_{\mathcal{P}} dj_3~
{\cal C}_{j_3}^{\alpha_3,\;w_3=w_1+w_2}\;
+\cdots \, ,\label{d+d-0}\\
{\cal D}^{\pm,\;w_1}_{j_1}\times\; {\cal
  C}^{\alpha_2,\;w_2}_{j_2}&=&\sum_{w=0}^1
\displaystyle  \int_{\mathcal{P}}
dj_3\;{\cal C}^{\alpha_3,\;w_3=w_1+w_2\pm w}_{j_3}
+ \sum_{j_3<-\frac 12}
{\cal D}^{\pm,\;w_3=w_1+w_2}_{j_3}\nonumber\\
&& + \sum_{j_3<-\frac 12}D^{\mp,w_3=w_1=w_2\pm 1}_{j_3}+\cdots \,,\label{dporc}\\
{\cal C}^{\alpha_1,\;w_1}_{j_1}\times
{\cal C}^{\alpha_2,\;w_2}_{j_2}
&=&\sum_{w=0}^1\sum_{j_3<-\frac
  12} \left ({\cal
    D}^{+,\;w_3=w_1+w_2-w}_{j_3}
+\;  {\cal
    D}^{-,\;w_3=w_1+w_2+w}_{j_3}\right )
\nonumber\\
&&+\sum_{w=-1}^1
\int_{\mathcal{P}} dj_3\;{\cal
  C}^{\alpha_3,\;w_3=w_1+w_2+w}_{j_3}+\cdots .\label{u}
\eea

In order to analyze these results, let us first restrict to the
spectral flow conserving contributions for
$w_i=0,~
 i=1,2.$ In this case, exactly the same
results were obtained in \cite{satoh} using the following prescription for
the OPE of $w=0$ primary fields
$\Phi^{j_1}_{m_1,\overline m_1}$ $\Phi^{j_2}_{m_2,\overline m_2}$
\footnote{~ See \cite{hs} for
 previous work involving highest-weight representations.}:

\begin{eqnarray}
\Phi^{ j_1}_{m_1,\overline m_1}(z_1,\overline z_1)\Phi^{j_2}_{m_2,\overline
  m_2} (z_2, \overline z_2){}^{~~\sim}_{z_1\rightarrow z_2}
\displaystyle\sum_{j_3}
  |z_{12}|^{-2\widetilde\Delta_{12}}
Q^{w=0}(j_i;m_i,\overline m_i)\Phi^{j_3}_{m_1+m_2,\overline
  m_1+\overline m_2}(z_2,
  \overline z_2),\label{opesatoh}
\end{eqnarray}
where $Q^{w=0}$ was obtained using the standard procedure, $i.e.$
multiplying both sides of (\ref{opesatoh}) by a fourth field in the
$w=0$ sector and taking 
expectation values.
 The formal symbol $\sum_{j_3}$ denotes
 integration over
 ${\cal
  D}_{j_3}^{\pm}$ and ${\cal C}_{j_3}^{\alpha_3}$,
  namely
\begin{eqnarray}
\sum_{j_3}=
\displaystyle\int_{{\cal P}^+}
d j_3 +
  \delta_{{\cal D}_{j_3}^{\pm}}\displaystyle\oint_{\cal C}dj_3 \, .\label{formal}
\end{eqnarray}
The integration over ${\cal P}^+$ stands for  summation over
$C_j^\alpha$.
The contour integral along
${\cal C}$
encloses the poles from ${\cal D}^{\pm}_{j_3}$
and $\delta_{{\cal
    D}_{j_3}^{\pm}}$ means that $j_3$ is picked up
from the poles in $Q^{w=0}$ by the contour ${\cal C}$
only when it
 belongs to a discrete representation. The range of $j_3$ is Re
 $j_3\le -\frac 12$ and Im $j_3\ge 0$, consistently with the argument
 which determined $Q^{w=0}$ because $\sum_{j_3}$ picks up only one
 term in (\ref{218}).
This prescription to deal with the $j-$dependent $m-$independent poles
 was shown  to be compatible
with the one suggested in \cite{tesch2} for the H$_3^+$ model.
 The strategy designed in (\ref{formal})  for the treatment of
$m-$dependent poles,  which were absent in \cite{tesch2}, aimed
 to reproducing
the classical tensor product of
 representations
of $SL(2, \mathbb R)$ in the limit $k\rightarrow \infty$ \footnote{~We thank
  Y. Satoh for comments on this point.}. This proposal for the OPE
includes in addition the requirement that poles with divergent residues
should not be picked up.

In this section, we have followed a different path.
 We have treated the $j-$ and
$m-$dependent poles alike.
However, although the
 equivalence between both prescriptions is not obvious a
 priori, we obtained the same
results for the OPE of unflowed primary fields \footnote{ More
 generally, it can be shown that a
generalization of the $ansatz$
(\ref{opesatoh}) for fields $\Phi^{j_1,w_1}_{m_1,\overline m_1}$
$\Phi^{j_2,w_2}_{m_2,\overline m_2}$, by adding the contributions from
terms proportional
to
$Q^{w=\pm 1}$ and replacing $\delta_{D_{j_3}^{\pm}}$
by $\delta_{D_{j_3}^{\pm,w_3}}$, leads to the same results
(\ref{1})$-$(\ref{u}).}.
Indeed, notice that
 poles in $Q^{w=0}$ at values of quantum numbers in $C_j^\alpha$
or ${\cal D}^{\pm}_{j_3}$
  would not
contribute to the OPE determined by (\ref{opec}) if
they do not cross the contour
${\cal P}$, unlike to (\ref{opesatoh}).
 On the other hand,
contributions from operators in other  representations, $i.e.$
neither in $C_j^\alpha$ nor  in ${\cal D}^{\pm}_{j_3}$,
could have appeared in (\ref{1})$-$(\ref{u}), but they did not. Moreover, by a
careful analysis of the analytic structure of $Q^{w=0}$ we
have shown  that there are no double poles,
so that the regularization proposed in \cite{satoh}
is not really necessary \footnote{ This is very important because the
double poles discussed in \cite{satoh}  would lead to inconsistencies in the analytic
 continuation of the OPE from H$_3^+$ that we have performed in this
 section. In particular, they would give divergent contributions to 
the OPE ${\cal D}^+_j\times{\cal D}^-_j$ and, in addition, this OPE would be
 incompatible with  ${\cal D}^-_j\times{\cal D}^+_j$, in
 contradiction with expectations from the symmetries of the function $W$.}.

In the case $w_1=w_2=0$, $k\rightarrow
\infty$, the $w-$conserving contributions to the OPE of
representations of the zero modes in (\ref{1})$-$(\ref{u})
reproduce the classical tensor products of
representations of $SL(2,\mathbb R)$ obtained in \cite{holman}.
Continuous series appear twice in the  product
of two continuous representations due to the existence of two
linearly independent Clebsh-Gordan coefficients. As noted in
\cite{satoh},
this is in agreement
with the fact that both terms $C^{12}$ and $C^{21}$ in (\ref{w})
contribute to $Q^{w=0}$ in the fusion  of two continuous series.
Moreover,
it was also observed that
the
analysis can be applied for finite $k$ without
modifications. The results are given by replacing ${\cal D}_j^\pm,
{\cal C}_j^{\alpha}$
in (\ref{1})$-$(\ref{u}) by the corresponding
affine representations
$\widehat{\cal D}_j^\pm, \widehat{\cal
  C}_j^\alpha$.
It is easy to see that this OPE of unflowed fields
in the
spectrum of the $SL(2,\mathbb R)$ WZNW model is not closed,
$i.e.$ it
gets contributions from discrete representations with $j_3<-\frac
{k-1}2$.
When  spectral flow is turned on,
 incorporating all the relevant representations
of the theory and the complete set of structure constants as we have
 done in this section, the OPE
still does not
close, namely there are contributions from discrete representations outside
the range (\ref{range}).  In particular, this feature of the OPE of
fields in discrete representations differs from the results in
\cite{mo3} where the factorization limit of the four-point function of
$w=0$ short strings was shown to be in accord with the Hilbert space
of the theory.

In the following section we will show that
assuming the OPE (\ref{1})$-$(\ref{u}) holds for states in
representations of the full current algebra, $i.e.$
replacing ${\cal D}_j^{\pm,w}, {\cal
  C}_j^{\alpha,w}$   by $\widehat{\cal D}_j^{\pm,w}, \widehat{\cal
  C}_j^{\alpha,w}$,
leads to inconsistencies unless  a truncation is performed.\\

\section{Truncation of the operator algebra and fusion rules}

The analysis of the previous section
 involved  primary operators and their spectral flow images.
Then, the OPE (\ref{1})-(\ref{u}) 
explicitly includes some descendant fields.
Assuming  the appearance of spectral flow images of primary states in the
 fusion rules
indicates that there are also contributions from
descendants not obtained by spectral flowing
 primaries but descendants with the same $J_0^3$
eigenvalue,  namely replacing ${\cal D}_j^{\pm,w}, {\cal C}_j^{\alpha,
 w}$ by $\hat{\cal D}_j^{\pm,w}, \hat{\cal C}_j^{\alpha,w}$ in the
 r.h.s. of (\ref{1})-(\ref{u}),
some interesting conclusions can be drawn.

For instance, consider the spectral flow non-preserving terms in
the OPE ${\cal D}^{-,w_1}_{j_1}\times{\cal
 D}_{j_2}^{-,w_2}$,  (\ref{1}). If they are extended to the affine series, using
 the spectral flow symmetry they may be identified as
\begin{equation}
\sum_{-\frac{k-1}2<\tilde j_3\leq
  j_1+j_2} \widehat{\cal D}^{+,\;w_3=w_1+w_2-1}_{-\frac k2-\tilde j_3}
\equiv\sum_{-\frac{k-1}2<j_3\leq
  j_1+j_2} \widehat{\cal D}^{-,\;w_3=w_1+w_2}_{j_3}\, .
\end{equation}
This reproduces the spectral flow conserving terms in the first sum in 
(\ref{1}).
However, there is an important difference:  here $j_3$ is
automatically restricted to
the region (\ref{unitary}).

Analogously, applying the spectral flow symmetry  to the discrete
series contributing to the OPE
$\displaystyle\left.{\cal D}^{+,w_1}_{j_1}\times{\cal D}_{j_2}^{-,w_2}\right|_{|w|= 1}$
in (\ref{d+td-}) leads to contributions from $\displaystyle
\sum_{j_2-j_1\leq j_3} {\widehat{\cal D}}^{-,\;w_3=w_1+w_2}_{j_3}$
 as well as from $\displaystyle\sum_{j_1-j_2\leq j_3} {\widehat{\cal D}}^
{+,\;w_3=w_1+w_2}_{j_3}$, which were found
among the spectral flow conserving terms
 with the extra condition $j_3<-\frac 12$.

In order to see further implications of the spectral flow symmetry on the OPE
 (\ref{1})-(\ref{u}), let us now consider operator products of descendants.
Take
 the OPE
 $\widehat{D}_{j_1}^{+,w_1=0}\otimes \widehat{D}_{j_2}^{-,w_2=1}$
\footnote{ We use the tensor product symbol $\otimes$ to denote the
 OPE of fields in representations of 
the current algebra, to distinguish it from that of
 highest/lowest-weight fields.}. Equation (\ref{d+d-0})
 gives  spectral flow conserving contributions from
$\widehat{\cal D}_{j_3}^{-,w_3=1}$, for certain $m_i,\overline m_i,
 i=1,2$, with $j_3$ verifying (\ref{unitary}).
Using the spectral flow symmetry,  one might infer that
the contributions from $\widehat{\cal D}^{+,w_3=0}_{j_3}$
to  the OPE $\widehat{\cal D}_{j_1}^{+,w_1=0}\otimes \widehat{\cal
 D}_{j_2}^{+,w_2=0}$ in (\ref{1}) would also be
 within the region (\ref{unitary}).
  On the contrary, we found 
 terms in $\widehat{\cal D}^{+,w_3=0}_{j_3}$ with $j_3<-\frac {k-1}{2}$.
Moreover, using the spectral flow symmetry again,
these terms can be identified with contributions from
 $\widehat{\cal D}^{-,w_3=1}_{j_3}$ with $j_3>-\frac
 12$ to the OPE $\widehat{\cal D}^{+,w_1=0}_{j_1}\otimes \widehat{\cal
 D}^{-,w_2=1}_{j_2}$, in contradiction with our previous result.

Similar puzzles are found identifying
 $\displaystyle\sum_{j_3<-\frac 12} \widehat{\cal
D}^{+,\;w_3=w_1+w_2-1}_{ j_3} =\displaystyle\sum_{-\frac{k-1}{2}<j_3} \widehat{\cal
D}^{-,\;w_3=w_1+w_2}_{ j_3}$  in (\ref{dporc}),
which gives some of the spectral flow conserving contributions.
 It is interesting to note that
only the states within the  region (\ref{unitary}) contribute in both cases,
explicitly $j_3=j_1+\alpha_2+n$, with $n\in \mathbb Z$ such
that $-\frac{k-1}2<j_3<-\frac 12$.
It is also important to stress the following observation. For given  $j_1,
m_1$ and $j_2,m_2$ the spectral flow conserving part of the OPE
 (\ref{dporc}) 
receives contributions
from states with $\tilde j_3, \tilde m_3$ verifying $\tilde j_3=\tilde
m_3+\tilde n_3$ with $\tilde n_3=0, 1,\cdots, \tilde n_{3}^{max}$,
$\tilde n_{3}^{max}$ being the maximum integer
such that $\tilde j_3<-\frac 12$.
On the other hand, the spectral flow non-conserving  terms
get contributions from  $j_3=- m_3+
n_3$
with $ n_3=0,1,2,\dots, n_{3}^{max}$ and here $ n_{3}^{max}$ is the
maximal non-negative integer such that $j_3<-\frac 12$. So,
identifying both series implies considering
 $\tilde j_3=-\frac k2- j_3$ and
now  $n_{3}^{max}$ (which is the same as before) has to be the
maximal non-negative integer for which $\tilde j_3>-\frac{k-1}{2}$.
There is just one operator appearing in both contributions to the OPE. It has
 $\tilde n_3=0$ in the former and
     $n_3=0$ in the latter.
This is a consequence of the relation
$\Phi_{m=\overline
m=-j}^{j,w=0}=\frac{\nu^{\frac k2-1}}{(k-2)}\frac{1}{B(-1-j')}
\Phi_{m'=\overline m'=j'}^{j',w'=1}$
with $j'=-\frac
k2-j$ \cite{mo3}. One
can check that the $w-$conserving three-point functions containing
$\Phi_{m=\overline m=-j}^{j,w=0}$ reduce to the $w-$non-conserving ones
involving $\Phi_{m'=\overline m'=j'}^{j',w'=1}$. This result can be
generalized for arbitrary $w$ sectors in the $m-$basis,
 $i.e.$ $\Phi_{m=\overline m=-j}^{j,w}\sim$
$\Phi_{m'=\overline m'=j'}^{j',w'=w+1}$ up to a regular
normalization for $j$ in the region (\ref{unitary}).
 For instance, one can reduce a spectral flow conserving three-point
function including $\Phi_{m=\overline m=-j}^{j,w}$ to a one unit
violating amplitude containing $\Phi_{m'=\overline m'=j'}^{j', w+1}$
using the identity
\bea
C(1+j_1,1+j_2,1+j_3)=\frac{\nu^{k-2}\gamma(k-2-j_{23})
\gamma(2-k-2j_1)C(k+j_1-1,1+j_2,1+j_3)}
{(k-2) \gamma(1+2j_1) \gamma(-N) \gamma(-j_{12}) \gamma(-j_{13})}\, ,
\label{Cidentity}
\eea
which is a consequence of the relation
$G(j)=(k-2)^{1+2j}\gamma(-j)G(j-k+2)$.\\

\bigskip

\centerline{\psfig{figure=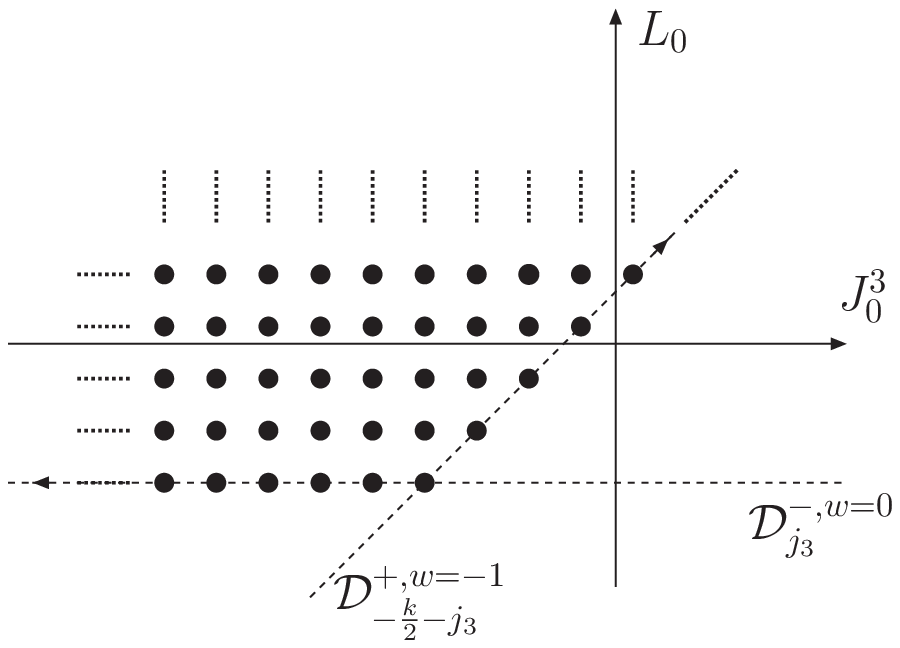,width=7cm}}
 {\footnotesize
{ Figure 5: Weight diagram of $\widehat{\cal D}^{-,w=0}_{j_3}$. 
The lines with arrows indicate the
states in ${\cal D}_{j_3}^{-,w=0}$ and ${\cal
 D}_{-\frac k2-j_3}^{+,w=-1}$. Consider
 a state in
$\widehat{\cal D}^{+,w=0}_{\tilde{j}}$, at level
 $\tilde{N}$ and
weight $\tilde{m}=-\tilde{j}+\tilde{n}$. It follows from
 (\ref{flowmap1}),
(\ref{flowmap2}) that after spectral flowing by ($-1$) unit, this
 state maps to a
state in $\widehat{\cal D}^{-,w=0}_{j}$, with  $j=-\frac k2-\tilde{j}$,
level $N=\tilde{n}$ and weight $m=j-n$, with $n=\tilde{N}$. For
 instance primary states
in $\widehat{\cal D}^{+,w=0}_{-\frac k2-j_3}$, denoted simply by
${\cal D}^{+,w=0}_{-\frac k2 -j_3}$, map to highest-weight states in
$\widehat{\cal D}^{-,w=0}_{j_3}$. So, only one state in ${\cal D}^{+,w=-1}_{-\frac
 k2-j_3}$
coincides with one in ${\cal D}^{-,w=0}_{j_3}$, namely that with
 $\tilde{n}=0$. }}
\bigskip

 The OPE (\ref{dporc}) was obtained for states in ${\cal C}^{\alpha,w}_j$ and
${\cal D}^{\pm, w}_j$.
When replacing operators in, say ${{\cal D}}^{-,w}_j$  by those in
${\widehat{\cal D}}^{-,w}_j$,
the latter can be  interpreted
as having been  obtained by performing
$w$ units of spectral flow on primaries of ${\widehat{\cal D}}^{-,w=0}_j$ or
$w-1$ units of spectral flow on primaries of ${\cal D}^{+}_{-\frac
k2-j}$, that is $w$ units of spectral flow from ${\cal D}^{+,w=-1}_{-\frac
k2-j}$, which in turn may be thought of as the highest-weight field
in ${\widehat{\cal D}}^{-,w=0}_j$ (see figure 5). Only the spectral flowed
primary  of highest-weight appears in both sets of contributions, $i.e.$
the one with $n_3 =\tilde n_3=0$. This behavior was observed in all other
cases,  namely, the same discrete series  arising in the OPE from
$Q^{w=0}$ can be also seen to arise from $Q^{w=1}$ or $Q^{w=-1}$, but only one operator 
appears
in both simultaneously.  

Thus, even if the
calculations involved operators in the series ${\cal D}_j^{\pm, w}$ and
${\cal C}_j^{\alpha, w}$,
we collect here the results for the fusion rules \footnote{ Actually,
the fusion rules for two representations determine the exact 
decomposition of their tensor products. These not only contain 
 information on the conformal families 
appearing in the r.h.s of the OPE, but also on their 
multiplicities. We shall not attempt to determine the latter here.} assuming 
$\Phi^{j_i,w_i}_{m_i,\overline
  m_i}(z_i,\overline z_i)\in\widehat{\cal D}_{j_i}^{\pm,w_i}$ or
$\widehat{\cal C}_{j_i}^{\alpha_i, w_i}$, $ i=1,2,3$.
Using the
  spectral flow symmetry to identify $\widehat{\cal D}_j^{-,w}=
\widehat{\cal D}_{-\frac k2-j}^{+,w- 1}$,
we obtain:

\noindent
\begin{enumerate}

\item ~~ $
\widehat{\cal D}^{+,\;w_1}_{j_1}
\otimes ~
\widehat{\cal D}^{+,\;w_2}_{j_2}
= \displaystyle\int_{\mathcal{P}} dj_3 ~ \widehat{\cal C}^{\alpha_3,\;
w_3=w_1+w_2+1}_{j_3}
 \oplus ~\sum_{-\frac{k-1}{2}<j_3\leq
  j_1+j_2} \widehat{\cal D}^{+,\;w_3=w_1+w_2}_{j_3}$ \label{ffr}

$~~~~~~~~~~~~~~~~~~~~~~~~~~~ \oplus ~\displaystyle
\sum_{j_1+j_2+\frac k2\leq j_3<-\frac
  12}  \widehat{\cal D}^{+,\;w_3=w_1+w_2+1}_{j_3},$

\noindent
\item $ ~~
\widehat{\cal D}^{+,\;w_1}_{j_1}\otimes ~
\widehat{\cal C}^{\alpha_2,\;w_2}_{j_2}
= \displaystyle
\sum_{-\frac{k-1}{2}<j_3<-\frac
  12} \widehat{\cal D}^{+,\;w_3=w_1+w_2}_{j_3}
\oplus ~
\sum_{w=0}^1\int_{\mathcal{P}} dj_3 ~
\widehat{\cal C}^{\alpha_3,\;w_3=w_1+
w_2+ w}_{j_3} ,$

\noindent
\item $ ~~ \widehat{\cal C}^{\alpha_1,\;w_1}_{j_1}
~ \otimes ~ \widehat{\cal C}^{\alpha_2,\;w_2}_{j_2}
= \displaystyle\sum_{w=-1}^0\sum_{-\frac{k-1}{2}<j_3<-\frac
  12} \widehat{\cal D}^{+,\;w_3=w_1+w_2+w}_{j_3}
\oplus ~~\displaystyle \sum_{w=-1}^1 \int_{\mathcal{P}} dj_3 ~
\widehat{\cal C}^{\alpha_3,\;w_3=w_1+w_2+w}_{j_3} .$

\end{enumerate}

We have truncated  the spin of the contributions from discrete
representations following the criterion that processes related through
the identity $\widehat{\cal D}_j^{+, w}\equiv \widehat{\cal
  D}_{-\frac k2-j}^{-, w+ 1}$ must be equal, $i.e.$
equivalent operator products should get the
same contributions.
Indeed,
one finds
contradictions unless the OPE is truncated to keep $j_3$ within
the region (\ref{unitary}). As we have seen through some examples,
extending the OPE (\ref{1})-(\ref{u}) to representations of the current algebra, 
discrepancies occur both when
comparing $w-$conserving with non-conserving contributions
as well as when comparing $w-$conserving
terms among themselves. So the truncation is imposed by self-consistency.

A strong argument
in support of the fusion rules 1.$-$3. is that
only  operators violating the
bound (\ref{unitary}) must be discarded.
Indeed, the cut  amounts to keeping just contributions from states
in the spectrum\footnote{ It is important to stress that
the truncation
is not discarding contributions
from the {\it microstates}
associated to the
$(j_1,j_2)-$dependent poles that were found in \cite{tesch2}. Only
$m-$dependent poles  which are absent in the $x-$basis present 
inconsistences with the spectral flow symmetry.}, $i.e.$
it implies that the operator algebra is closed on the Hilbert
space of the theory. However, the spectrum involves
irreducible representations and there are no singular vectors to
decouple states like in $SU(2)$  \cite{zf}
\footnote{~The spectral flow operators $\Phi^{-\frac k2}_{\pm\frac
k2,\pm\frac k2}$ have null descendants. Even
though they are excluded from the range (\ref{unitary}) they are necessary
auxiliary fields to construct the states in spectral flow
representations. Although the  physical mechanism is not clear to us,
these operators might play a role in the decoupling.}. We do not have an understanding of
the physical process determining  the
truncation.
Moreover, the cut cannot be directly implemented in the analysis
performed in the previous section because it would break analyticity.
Therefore, either the prescription
(\ref{opec}) must be modified to be consistent with the spectral
flow symmetry or there is a yet to be discovered physical mechanism
decoupling states. In other words,
the OPE in the H$_3^+$  and the
AdS$_3$ WZNW models do not seem to be just related by analytic
continuation, at least not in the way we have implemented here.

Nevertheless, the results listed in  items 1.$-$3. above 
 are supported by several consistency checks.
First, the limit $k\rightarrow \infty$ contains the classical tensor
 products of representations of $SL(2,\mathbb R)$
 \cite{holman} when restricted to $w=0$ fields.
Second, as mentioned in the previous paragraph, 
once the OPE is truncated  to keep only  contributions from
the spectrum, one can verify full consistency. In
 particular, the OPE $\widehat{\cal D}_{j_1}^{+,w_1}\otimes
\widehat{\cal D}^{+,w_2}_{j_2}$ is consistent with the results in \cite{mo3}
(see the discussion in appendix A.2).
Finally, based on
the spectral flow selection rules (\ref{cc}) and (\ref{dd}), the
following alternative analysis can be performed.
Let us
consider, for instance, the operator product
$\widehat{\cal D}_{j_1}^{+,w_1}\otimes
\widehat{\cal D}^{+,w_2}_{j_2}$.
Applying
equation (\ref{dd}) to
correlators involving three discrete states in $\widehat{\cal D}_{j}^{+,w}$
requires either
 $i) ~w_3=-w_1-w_2-1$
or $ii) ~w_3=-w_1-w_2-2$. Therefore, together with $m$ conservation,
$i)$ implies
 that the
three-point function
$<\widehat{\cal D}_{j_1}^{+,w_1}
\widehat{\cal D}^{+,w_2}_{j_2}\widehat{\cal
  D}_{j_3}^{+,w_3=-w_1-w_2-1}> $
will not vanish
as long as the OPE
$\widehat{\cal D}_{j_1}^{+,w_1}\otimes
\widehat{\cal D}^{+,w_2}_{j_2}$ contains a state in
$\widehat{\cal D}_{j_3}^{-,w=w_1+w_2+1}$, which is equivalent to
$\widehat{\cal D}_{\tilde j_3}^{+,w=w_1+w_2}$. Indeed, this contribution appeared above.
Similarly, $ii)$ implies that  in order for $<\widehat{\cal D}_{j_1}^{+,w_1}
\widehat{\cal D}^{+,w_2}_{j_2}\widehat{\cal
 D}_{j_3}^{+,w_3=-w_1-w_2-2}>$ to be non-vanishing, 
the OPE
$\widehat{\cal D}_{j_1}^{+,w_1}\otimes
\widehat{\cal D}^{+,w_2}_{j_2}$ must have contributions from
$\widehat{\cal D}_{j_3}^{-,w_3=w_1+w_2+2}\equiv \widehat{\cal
   D}_{\tilde j_3}^{+,
w_3=w_1+w_2+1}$,
which in fact
were found. Finally, when the third state involved in the three-point
function is in the series $\widehat{\cal C}_{j_3}^{\alpha_3,w_3}$,
equation (\ref{cc})
leaves
only one possibility, namely $w_3=-w_1-w_2-1$, and thus the OPE must
include terms in $\widehat{\cal C}_{j_3}^{\alpha_3, w_3=w_1+w_2+1}$, which actually
appear in
the list above.
Although this analysis based on the spectral flow selection rules
 does not allow to determine either the range of $j_3-$values or the OPE coefficients,
it is easy to check that
the series content in 1.$-$3. is indeed completely reproduced in this way.

As mentioned in the previous section, in principle $w=\pm 2$ 
three-point functions should have been considered. However, the
contributions from these terms are already contained in our
results. If they gave contributions from discrete representations
outside the spectrum, they should be truncated since the equivalent
terms listed above do not include them. Contributions from operators
in $\widehat{\cal D}_{j_3}^{-,w_3=w_1+w_2+2}$ can only appear in case
1., namely $\widehat{\cal D}_{j_1}^{+,w_1}\otimes
\widehat{\cal D}^{+,w_2}_{j_2}$,
 for $j_3=-k-j_1-j_2-n$. These correspond to the terms denoted as
Poles$_2$ in \cite{mo3}, where they could not be interpreted in terms
of physical string states and were then truncated. See appendix A.2 for a
detailed discussion.

In conclusion,
the results presented in this section are in agreement with the
spectral flow selection pattern (\ref{cc})-(\ref{dd}), they are
consistent with the results in \cite{mo3}  and
determine the closure of the operator algebra
when properly treating the spectral flow symmetry. 
The full consistency of the OPE  should follow
from a proof of factorization and crossing symmetry of the four-point
functions, but closed expressions for these amplitudes 
are not known,
even in the simpler H$_3^+$ model. 
In order to make some preliminary
progress in this direction,
in the next section we discuss certain properties of the factorization of
four-point
amplitudes  involving states in
different representations of the $SL(2,\mathbb R)$ WZNW model, constructed
along the lines in
\cite{tesch2}.

\section{Comments on the factorization of  four-point functions}

Although a complete description of the
contributions of descendant operators 
is not available to complete the bootstrap program, in this section we  
 display some
interesting properties of the amplitudes that can be useful to
achieve a resolution of the theory.
We first summarize known results on the $s-$channel
factorization of four-point functions in the H$_3^+$ model and 
show that an alternative expression can be written in the
AdS$_3$ WZNW model if the correlators in both models are related
through analytic continuation. Then,
we perform a qualitative study of the contributions of primaries and flowed
primaries in the intermediate channels of the amplitudes and finally, we
discuss the
consistency of the factorization with the spectral flow selection rules.

A decomposition of the
four-point function in the Euclidean model was
worked out
in \cite{tesch1, tesch2}
using the OPE (\ref{ope})
for  pairs of primary operators $\Phi_{j_1}\Phi_{j_2}$ and
$\Phi_{j_3}\Phi_{j_4}$.
The $s-$channel factorization was
written as follows

\begin{eqnarray}
\left\langle \Phi_{j_1}(x_1|z_1)\Phi_{j_2}(x_2|z_2)
\Phi_{j_3}(x_3|z_3)\Phi_{j_4}(x_4|z_4)
\right\rangle =
|z_{34}|^{2(\tilde\Delta_2+\tilde\Delta_1-\tilde\Delta_4-\tilde\Delta_3)}
|z_{14}|^{2(\tilde\Delta_2+\tilde\Delta_3-
\tilde\Delta_4-\tilde\Delta_1)}\nonumber\\
\times~|z_{24}|^{-4\tilde\Delta_2}
~|z_{13}|^{2(\tilde\Delta_4-\tilde\Delta_1-
\tilde\Delta_2-\tilde\Delta_3)}
\int_{{\cal P}^+}dj ~
{\cal A}(j_i, j)~
{\cal G}_j(j_i,z,\overline
z,x_i,\overline x_i)~|z|^{2(
  \Delta_j-\Delta_{1}-\Delta_{2})} .
\label{4pt}
\end{eqnarray}
Here
\begin{equation}
{\cal A}(j_i,j)=C(-j_1,
-j_2,-j)B(-j-1)
C(-j,-j_3,-j_4)
\end{equation}
and
\begin{equation}
{\cal G}_j(j_i,z,\overline z,x_i,\overline x_i)=\sum_{n,\overline
  n =0}^{\infty}z^n\overline z^{\overline
  n}D^{(n)}_{x,j}(j_i,x_i)
\overline D^{\overline n}_{\overline x,
  j}(j_i,\overline x_i)G_j(j_i,x_i, \overline x_i)\, ,
\end{equation}
where $D^{(n)}_{x,j}(j_i,x_i)$ are differential operators containing
the contributions from intermediate descendant states and
\begin{eqnarray}
 G_j(j_i, x_i, \overline x_i) & = &|x_{12}|^{2(j_{1}+j_2-j)}
|x_{34}|^{2(j_{3}+j_4-j)}
\int  d^2x d^2x'
|x_{1}-x|^{2(j_{1}+j-j_2)}
|x_{2}-x|^{2(j_{2}+j-j_1)}\nonumber\\
&&\times ~
|x_3-x'|^{2(j_{3}+j-j_4)}|x_4-x'|^{2(j_{4}+j-j_3)}|x-x'|^{-4j-4}\, ,\label{gj}
\end{eqnarray}
which may be rewritten as
\begin{eqnarray}
G_j(j_i, x_i, \overline x_i) & = &\frac{\pi^2}{(2j+1)^2}|x_{34}|^{2(j_{4}+j_3-j_2-j_1)}
|x_{24}|^{4j_{2}}|x_{14}|^{2(j_{4}+j_1-j_2-j_3)}|x_{13}|^{2(j_{3}+j_2+j_1-j_4)}
\nonumber\\
&\times &\left\{|F_j(j_i,x)|^2+
\frac{\gamma(1+j+j_4-j_3)\gamma(1+j+j_3-j_4)}{\gamma(2j+1)
\gamma(j_1-j_2-j)\gamma(j_2-j_1-j)}
|F_{-1-j}(j_i,x)|^2\right\}\, ,\nonumber
\end{eqnarray}
with $F_j(j_i, x)\equiv x^{j_1+j_2-j} {}_{2}F_1(j_1-j_2-j,j_4-j_3-j;-2j;x)$
and $x=\frac {x_{12}x_{34}}{x_{13}x_{24}}$.

The properties of (\ref{4pt}) under $j\rightarrow -1-j$ allow to extend
the integration contour from ${\cal P}^+$ to the full axis ${\cal
  P}=-\frac 12+i{\mathbb R}$ and rewrite it in a holomorphically factorized form.
Crossing symmetry 
follows from similar properties of a
five-point function in Liouville theory and it amounts to establishing
the consistency of the H$_3^+$ WZNW model  \cite{tesch3}.

Expression (\ref{4pt}) is valid for external states
$\Phi_{j_1},\Phi_{j_2}$
in the range (\ref{range}) and similarly
for $\Phi_{j_3}, \Phi_{j_4}$. In particular, it holds for operators
 in continuous representations of the $SL(2,\mathbb R)$ WZNW model.
The analytic continuation to other values of
$j_i$ was performed in
\cite{mo3}. In this process, some poles in the integrand cross the
integration contour and the four-point function is defined as
(\ref{4pt}) plus the contributions of all these poles.
This procedure allowed to analyze the factorization of 
four-point functions of $w=0$ short
strings in the boundary conformal field theory, obtained from primary states in discrete
representations ${\cal D}_j^{ w=0}\otimes {\cal D}_j^{
  w=0}$, by integrating over the world-sheet moduli.
It is important to stress that the aim in \cite{mo3} was to study
the factorization in the boundary conformal field theory with
coordinates  $x_i, \overline x_i$, so
the $x-$basis was found convenient. The conformal blocks were expanded
in powers of the cross ratios $x, \overline x$ and then integrated
over the worldsheet coordinates
$z,\overline z$. 
To study the factorization in the $SL(2,\mathbb R)$
WZNW model instead, we expand the conformal blocks in powers of $z,
\overline z$, and in order to consider the various sectors, we find
convenient to translate (\ref{4pt}) to the $m-$basis. 

To this purpose, one can
verify that the integral over $j$ commutes with the integrals over
$x_i,\overline x_i,\,i=1,\dots,4$ and that it is regular for
$j_{21}^{\pm}$ and $j_{43}^{\pm}$ in the range (\ref{range}) and for all of
$|m|,|\overline m|, |m_i|,|\overline m_i|<\frac12$, where we have
introduced $m=m_1+m_2=-m_3-m_4$, $\overline m=\overline
m_1+\overline m_2=-\overline m_3 -\overline m_4$. Integrating in
addition over $x$ and $x'$ in (\ref{gj}), we get
\begin{eqnarray}
\left\langle \Phi^{j_1}_{m_1,\overline m_1}\Phi^{j_2}_{m_2,\overline m_2}
\Phi^{j_3}_{m_3,\overline m_3}\Phi^{j_4}_{m_4,\overline m_4}
\right\rangle  =
|z_{34}|^{2(\tilde\Delta_2+\tilde\Delta_1-
\tilde\Delta_4-\tilde\Delta_3)}
|z_{14}|^{2(\tilde\Delta_2+\tilde\Delta_3-
\tilde\Delta_4-\tilde\Delta_1)}|z_{24}|^{-4\tilde\Delta_2}\nonumber\\
~~~~~~~~\times ~ |z_{13}|^{2(\tilde\Delta_4-\tilde\Delta_1-
\tilde\Delta_2-\tilde\Delta_3)}
\int_{{\cal P}^+}dj ~
{\mathbb A}_j^{w=0}(j_i;m_i,\overline m_i)~
|z|^{2(
  \tilde\Delta_j-\tilde\Delta_{1}-\tilde\Delta_{2})}+\cdots\, ,\label{4ptm}
\end{eqnarray} 
where
\bea 
{\mathbb A}_j^{w=0}(j_i;m_i,\overline m_i)& = &
\delta^{(2)}(m_1+\dots+m_4)~C(1+j_1,1+j_2,1+j)~W\left[\begin{matrix}
j_1\,,j_2,\,j\cr m_1,m_2,-m\cr\end{matrix}\right]\cr &&\times ~
\frac{1}{B(-1-j)~c^{-1-j}_{m,\overline m}} ~C(1+j_3,1+j_4,1+j)~ W\left[\begin{matrix}
j_3\,,j_4,\,j\cr m_3,m_4,m\cr\end{matrix}\right].\label{A4pt1}
\eea

An alternative representation of (\ref{A4pt1}) was found in
\cite{mn} in terms of higher generalized hypergeometric functions
$_4F_3$.
This new
identity among hypergeometric functions is an interesting
by-product of the present result.

The dots in (\ref{4ptm}) refer to higher
powers of $z, \overline z$ corresponding to the integration of terms
of the form
$\mathbb A_j^{N,w=0}|z|^{2(
  \Delta^{(N)}_j-\tilde\Delta_{1}-\tilde\Delta_{2})}$, where
$\mathbb A_j^{N,w=0},\;N=1,2,3,\dots$
  stand for  contributions from
descendant operators at level $N$ with conformal weights $
  \Delta^{(N)}_j=
  \tilde\Delta_j+N$.

Notice that the symmetry under $j\leftrightarrow-1-j$ in (\ref{A4pt1}),
which can be easily checked by using the identity (\ref{eqsat}), allows  to
extend the integral to the full axis ${\cal P} = -\frac{1}{2}+i\mathbb{R}$.

Given that
correlation functions in the $SL(2,\mathbb R)$ WZNW model in the $m-$basis depend on the
sum of $w_i$ numbers,
 except for the powers of the coordinates $z_i, \overline z_i$, if the
 Lorentzian and Euclidean theories are simply related by analytic continuation,
this result should
hold, in particular, for states in continuous representations 
 in arbitrary spectral flow sectors (with $|m_i|,|\overline m_i|, |m|
<\frac 12$), as long as
$\sum_iw_i=0$, $i.e.$
\begin{eqnarray}
\left\langle \Phi^{j_1, w_1}_{m_1,\overline m_1}\Phi^{j_2,w_2}_{m_2,\overline m_2}
\Phi^{j_3,w_3}_{m_3,\overline m_3}\Phi^{j_4,w_4}_{m_4,\overline m_4}
\right\rangle _{\sum_{i=1}^4w_i=0} ~= ~
z_{34}^{\Delta_2+\Delta_1-
\Delta_4-\Delta_3}z_{14}^{\Delta_2+\Delta_3-
\Delta_4-\Delta_1}z_{13}^{\Delta_4-\Delta_1-
\Delta_2-\Delta_3}
\nonumber\\
\times ~ z_{24}^{-2\Delta_2}
\times ~c.c.~\times
\int_{\cal P} dj ~
{\mathbb A}_j^{w=0}(j_i;m_i,\overline m_i)~
z^{\Delta_j-\Delta_{1}-\Delta_{2}}\overline z^{\overline
\Delta_j-\overline\Delta_{1}-\overline\Delta_{2}}
+\cdots \, ,
\label{final4}
\end{eqnarray}
where $\Delta_j=-\frac {j(j+1)}{k-2}-m(w_1+w_2)-\frac k4(w_1+w_2)^2$
 and
 $c.c.$ stands for the obvious antiholomorphic $\overline
z_i-$dependence.
For other values of
$j_1, \cdots ,j_4$, $m_1,\dots,\overline m_4$ the integral
may diverge and must be
defined by analytic continuation.

That a generic $w-$conserving four-point function involving primaries or
highest/lowest-weight states in 
${\cal C}_j^{\alpha,w}$ or ${\cal
  D}_j^{\pm,w}$ 
should factorize
as in (\ref{final4}), if the amplitude with four $w=0$ states is 
given by (\ref{4ptm}), can be deduced from
the  relation \cite{ribault}:
\begin{equation}
\displaystyle\left\langle \prod_{i=1}^n \Phi_{m_i,\overline m_i}^{j_i,w_i}
(z_i,\overline z_i)\right\rangle_{\sum_{i=1}^n w_i=0}=\kappa
\overline{\kappa}\left\langle \prod_{i=1}^n
\Phi_{m_i,\overline m_i}^{j_i,\widetilde{w}_i=0}(z_i,\overline
z_i)\right\rangle,
\end{equation}
where $\displaystyle\kappa=\prod_{i<j}z_{ij}^{-w_i m_j-w_j m_i-\frac
  k2 w_i w_j}$,
$\displaystyle\overline\kappa=\prod_{i<j}z_{ij}^{-w_i \overline
m_j-w_j \overline m_i-\frac k2 w_i w_j}$,
after Taylor expanding around $z=0$ the r.h.s. of the following identity:
\bea
&&\kappa~~
z_{34}^{\tilde\Delta_2+\tilde\Delta_1-\tilde\Delta_4-\tilde\Delta_3}
z_{14}^{\tilde\Delta_2+\tilde\Delta_3-\tilde\Delta_4-\tilde\Delta_1}
z_{24}^{-2\tilde\Delta_2} z_{13}^{\tilde\Delta_4-\tilde\Delta_1-
\tilde\Delta_2-\tilde\Delta_3}z^{\tilde\Delta_j-\tilde\Delta_{1}-\tilde\Delta_{2}}=\cr
&& z_{34}^{\Delta_2+\Delta_1-\Delta_4-\Delta_3}
z_{14}^{\Delta_2+\Delta_3-\Delta_4-\Delta_1}
z_{13}^{\Delta_4-\Delta_1-
\Delta_2-\Delta_3}z_{24}^{-2\Delta_2}
z^{\Delta_j-\Delta_{1}-\Delta_{2}}\left(1-z\right)^{-m_2w_3-m_3w_2-
\frac k2w_2w_3}.\label{kappa}\nonumber\\
\eea

The conclusion is that, if the H$_3^+$ and AdS$_3$ models are simply related by
analytic continuation, then 
(\ref{final4}) and its analytic continuation should hold for
generic $w-$conserving four-point functions of fields in 
${\cal C}_j^{\alpha,w}$ or ${\cal
  D}_j^{\pm,w}$ \footnote{See appendix A.3 for an alternative discussion
directly in the $m-$basis, independent of the $x-$basis.}.
However, expression  (\ref{final4}) appears to be in 
contradiction with the
factorization $ansatz$ 
and the OPE found in section 3 for the $SL(2,\mathbb R)$ WZNW model, 
because it seems to contain
 just $w-$conserving channels.
Actually, directly applying the
factorization $ansatz$ based on the OPE (\ref{opec}) would give the
following expression for both $w-$conserving and violating four-point functions:
\begin{equation}
\left\langle \Phi^{j_1, w_1}_{m_1,\overline m_1}\Phi^{j_2,w_2}_{m_2,\overline m_2}
\Phi^{j_3,w_3}_{m_3,\overline m_3}\Phi^{j_4,w_4}_{m_4,\overline m_4}
\right\rangle  ~\sim ~
z_{34}^{\Delta_2+\Delta_1-
\Delta_4-\Delta_3}z_{14}^{\Delta_2+\Delta_3-
\Delta_4-\Delta_1}z_{13}^{\Delta_4-\Delta_1-
\Delta_2-\Delta_3} z_{24}^{-2\Delta_2}\times c.c.
\nonumber
\end{equation}
\begin{equation}
~~\times~\delta^2(\sum_{i=1}^4m_i+\frac k2w_i) \sum_{w=-1}^1\int_{\cal P} dj ~
Q^wQ^{-w-\sum_{i=1}^4w_i}B(-1-j)c^{-1-j}_{m,\overline m}
z^{\Delta_j-\Delta_{1}-\Delta_{2}}\overline z^{\overline
\Delta_j-\overline\Delta_{1}-\overline\Delta_{2}}
+\cdots
\label{f4pt}
\end{equation}
with $m=m_1+m_2-\frac k2w = -m_3-m_4-\frac k2w$,
$\overline m=\overline m_1+\overline m_2-\frac k2w = -\overline
m_3-\overline m_4-\frac k2w$ and $\Delta_j=-\frac
{j(j+1)}{k-2}-m(w_1+w_2+w)-\frac k4(w_1+w_2+w)^2$  (similarly for
$\overline \Delta_j$).
Actually,
in the $m-$basis, the starting point for the $w-$conserving four-point
function would have been 
(\ref{final4}) plus an analogous contribution involving one unit spectral 
flow three-point functions, $i.e.$
(\ref{final4}) rewritten in terms of ${\mathbb A}_j^{w=1}$
or ${\mathbb A}_j^{w=-1}$ instead of ${\mathbb A}_j^{w=0}$, where
\bea
{\mathbb A}_j^{w=\pm 1}(j_i;m_i,\overline m_i) &=&
\delta^{(2)}(\sum_{i=1}^4m_i)~\frac{\widetilde
C(1+j_1,1+j_2,1+j)}{\gamma(j_1+j_2+j+3-
\frac k2)}
\widetilde
W\left[\begin{matrix}
j_1~,~~j_2~~,~j\cr \mp m_1, \mp m_2, \pm
m\cr\end{matrix}\right]\cr
&\times& \frac{1}{B(-1-j)c^{-1-j}_{m,\overline m}}
\frac{\widetilde C(1+j_3,1+j_4,1+j)}{\gamma(j_3+j_4+j+3-\frac k2)}
\widetilde W\left[\begin{matrix}
j_3~,~~j_4~~,~j\cr \pm m_3, \pm m_4, \pm m \cr\end{matrix}\right]
.\label{Aw1}
\eea
But if correlation functions in this model
 are to be obtained from those in the H$_3^+$ model
\cite{mo3, tesch1, tesch2, hs, satoh,  gk},
spectral flow conserving and non-conserving channels should give the same
result for the $w-$conserving four-point functions. 
This does not imply that $\mathbb A_j^{w=0}$ and $\mathbb A_j^{w=\pm1}$
carry the same amount of information \footnote{ In other words, both
 expressions seem to give the same contribution in $w-$conserving four-point
functions. However one cannot always use either one of them. In
particular, this is not expected to hold for $w-$violating amplitudes.}. In
general, if both expressions for the four-point functions 
were equivalent, one would expect that part of the
information in $\mathbb A_j^{w=0}$ were contained in $\mathbb
A_j^{w=\pm1}$ and the rest in the
contributions from descendants in $\mathbb A_j^{N,w=\pm1}$.

A proof of this statement would require making  explicit  the
higher order terms and possibly some contour manipulations,
 which we shall not attempt. Nevertheless there are several
 indications supporting this claim.
A similar proposition
was advanced in \cite{ribault} for the H$_3^+$ model and some
evidence was given that these possibilities might not be exclusive,
depending on which correlator the OPE is inserted in.
Furthermore,
$w=1$ long strings were
found in the $s-$channel factorization of the four-point amplitude
of $w=0$ short strings in \cite{mo3} starting from the holomorphically
factorized expression for (\ref{4pt}), rewriting the integrand and
moving the integration contour. Moreover,
in the $m-$basis, spectral flow non-conserving channels can be seen to
appear naturally from
 (\ref{final4}) in certain special cases, as we now show.

Identities
among different expansions of four-point functions containing at least one field
in discrete representations can be generated using the spectral flow symmetry.
In particular,  $w-$conserving four-point functions
 involving the fields $\Phi_{m_1=\overline m_1=-j_1}^{j_1,w_1}$ and  
$\Phi_{m_3=\overline m_3=j_3}^{j_3,w_3}$ coincide (up to
 $B(j_1),B(j_3)$ factors) with the 
 $w-$conserving amplitudes involving
$\Phi_{m'_1=\overline m'_1=j'_1}^{j'_1=-\frac k2-j_1,w'_1=w_1+1}$ and  
$\Phi_{m'_3=\overline m'_3=-j'_3}^{j'_3=-\frac k2-j_3,w'_3=w_3-1}$
\footnote{ This is a consequence of the identities discussed in the
  paragraph containing equation (\ref{Cidentity}) in the previous section.}. This allows
to expand the four-point amplitude in two alternative ways, namely
\bea
\int_{{\cal P}}dj ~ {\mathbb
A}_j^{w=0}(j_1,j_2,j_3,j_4;m_1,\dots,\overline m_3,\overline m_4)~ z^{
  \Delta(j)-\Delta(j_1)-\Delta(j_2)} \overline z^{\overline
  \Delta(j)-\overline\Delta(j_1)-\overline\Delta(j_2)}
+\cdots\label{factAw0}
\eea
or
\bea
\beta_{1,3}\int_{{\cal P}}dj ~
{\mathbb A}_j^{w=0}(j'_1,j_2,j'_3,j_4;m'_1,\dots,\overline m'_3,\overline m_4)~
  z^{\Delta'(j)-\Delta(j'_1)-\Delta(j_2)}
\overline z^{\overline\Delta'(j)-\overline\Delta(j'_1)-\overline\Delta(j_2)}
+ \cdots ,  \label{A'}
\eea 
where $\beta_{1,3}\equiv\frac{B(-1-j_3)}{B(-1-j'_1)}$ and the dots refer  
to  contributions from descendants and, in addition, to
residues at poles in $\mathbb A_j^{w=0}$ crossing ${\cal P}$
after analytic continuation of $j_i$ ($i=1,3$ and eventually $2,4$) to the
region (\ref{unitary}). Explicitly, ${\mathbb
A}_j^{w=0}(j'_1,j_2,j'_3,j_4;m'_1,\dots,\overline m'_3,\overline m_4)$ is 
given by 
\bea 
&&C(1+j'_1,1+j_2,1+j)C(1+j'_3,1+j_4,1+j)
\frac{\pi^3\gamma(2+2j)}{B(-1-j)}
\frac{\gamma(j-j'_1-j_2)\gamma(j_2-j'_1-j)
}{\gamma(2+j'_1+j_2+j)\gamma(-2j'_1)}\cr\cr &&\times
\frac{\gamma(j-j'_3-j_4)\gamma(j_4-j'_3-j)
}{\gamma(2+j'_3+j_4+j)\gamma(-2j'_3)}
\frac{\Gamma(1+j_2-m_2)\Gamma(1+j_4+\overline 
m_4)}{\Gamma(-j_2+\overline m_2) \Gamma(-j_4-m_4)}
\frac{\Gamma(-j-\overline m)
\Gamma(1+j-m)}{\Gamma(1+j+m)\Gamma(-j+\overline m)} .\nonumber
\eea

Using  (\ref{Cidentity}) and rewriting this expression in terms of $j_i,m_i$, the
following equivalence can be shown
\bea
(\ref{A'})=\int_{{\cal P}}dj ~
{\mathbb A}_j^{w=1}(j_1,j_2,j_3,j_4;m_1,\dots,\overline m_3,\overline m_4)~
z^{\Delta(j)-\Delta(j_1)-\Delta(j_2)}\overline 
z^{\overline\Delta(j)-\overline\Delta(j_1)-\overline \Delta(j_2)}+
\cdots.\label{factAw1}
\eea

Notice that not only the coefficient $\mathbb A_j^{w=1}$  but also the
$z_i,\overline z_i$ dependence are as expected.
In fact, $\Delta(j'_1)=\tilde\Delta(j'_1)-m'_1 w'_1-\frac k4 w'_1{}^2=$
$\tilde\Delta(j_1)-m_1w_1-\frac k4w_1^2=\Delta(j_1)$ and
$\Delta'(j)=\tilde\Delta(j)-(m'_1+m_2)(w'_1+w_2)-\frac
k4(w'_1+w_2)^2=$
$\tilde\Delta(j_1)-mw-\frac k4w^2=\Delta(j_1)$,
where $m=m_1+m_2-\frac k2$ and $w=w_1+w_2+1$. Therefore, we have seen
in a particular example that spectral flow conserving and violating
channels can give the same result for four-point functions.
This is a nontrivial result showing that 
the spectral flow 
symmetry allows to exhibit
$w-$non-conserving channels that are not equivalent to other $w-$conserving 
ones in expressions constructed as sums over
 $w-$conserving exchanges. 

In appendix A.3 we  show that the terms explicitly displayed in both
(\ref{final4}) and (\ref{factAw1}) are 
solutions of the Knizhnik-Zamolodchikov (KZ) equations. 
However, these equations do not give 
 enough information  to 
confirm that the full expressions  (\ref{final4}) and (\ref{factAw1}) 
are equivalent.

The factorization of  four-point functions 
reproduces  the field
content of the OPE. 
Therefore, the truncation imposed on the operator algebra
by the spectral flow symmetry must be realized in physical amplitudes. Again,
to confirm this would
require  more information on the contributions from descendant fields
and studying
crossing symmetry.
Here, we just illustrate this point with one example.
Take for instance the following four-point function \footnote{ Here,
  as in the previous section, we
denote the states by the representations they belong to and we omit
the antiholomorphic part for short.}:
\begin{equation}
\left\langle{\cal D}^{+,w_1=0}_{j_1}{\cal D}^{+,w_2=-1}_{j_2}
{\cal D}^{-,w_3=0}_{j_3}{\cal D}^{-,w_4=-1}_{j_4}
\right\rangle\, ,\label{4pte}
\end{equation}
in the particular case with $n_i=0,\forall i$ (where $m_i=\pm j_i\mp
n_i$) and
$j_1+j_2=j_3+j_4<-\frac{k-1}{2}$. The OPE (\ref{1}) implies one
intermediate state
in the $s-$channel in ${\cal D}^{+,w=-1}_j$, with $j=j_1+j_2=-m$ as
well as exchanges of
states in
${\cal D}^{+,w=0}_j$  if $j_1+j_2=j_3+j_4<-\frac{k+1}2$ with
$j=j_1+j_2+\frac k2+n$, $n=0,1,2,\dots$ such that $j<-\frac12$, and
also of continuous states in ${\cal C}_j^{\alpha, w=0}$. The unique state found in ${\cal
   D}^{+,w=-1}_j$
is equivalent to the highest-weight state in ${\cal D}^{-,w=0}_{\tilde j}$
with
$\tilde j=-\frac k2-j>-\frac 12$.

This four-point
function must
coincide with the following one:
\begin{equation}
\left\langle{\cal D}^{+,w_1=0}_{j_1}{\cal D}^{-,w_2=0}_{\tilde j_2}
{\cal D}^{-,w_3=0}_{j_3}{\cal D}^{+,w_4=0}_{\tilde j_4}
\right\rangle\, ,
\end{equation}
where as usual $\tilde j_i=-\frac k2-j_i$ (notice that this holds
without ``hats"
because $n_i=0, \forall i$). Now $\tilde j_2-j_1=\tilde
 j_4-j_3>-\frac 12$.
Therefore, (\ref{d+d-0}) implies that only states from ${\cal C}_j^{\alpha, w=0}$
as well as  from
${\cal D}^{+,w=0}_j$ with $j=j_1-\tilde j_2+n=j_1+j_2+\frac k2+n$
 propagate in the
intermediate $s-$channel, the latter  requiring the extra condition
 $\tilde j_2-j_1=\tilde
j_4-j_3>\frac 12$, $i.e.$
 $j_1+j_2=j_3+j_4<-\frac{k+1}{2}$. The
important remark is that no intermediate states from ${\cal D}^{-,w=0}_{\tilde
 j}$ appear in the
factorization.
 This behavior was discussed in the previous section
when studying the consequences of the spectral flow symmetry on the OPE. However, we
 have considered
this case carefully here because it explicitly displays the fact that
 the same
four-point function
factorizes in two different ways and the unique difference is an extra state
 violating the bounds (\ref{unitary}). Recall that we are only
 considering primaries and their spectral flow images.
We expect that some consistency requirements, such as
crossing symmetry, will 
automatically realize
the OPE displayed in the previous section   in 
physical amplitudes.

An indication in favor of the bootstrap approach to
this non-rational CFT is that the expressions
reproduce the spectral flow selection rules (\ref{cc}) and
(\ref{dd}) for four-point functions in different sectors. Indeed,
let us analyze this feature in a
four-point function involving only external discrete states
or their spectral flow images.
The  bounds (\ref{dd}) require
$-3\le \sum_{i=1}^4 w_i \le -1$,
in agreement  with the  factorization of this
amplitude in any channel.
Indeed, consider for instance
\begin{equation}
\left\langle{\widehat{\cal D}}^{+,w_1}_{j_1}{\widehat{\cal D}}^{+,w_2}_{j_2}
{\widehat{\cal D}}^{+,w_3}_{j_3}{\widehat{\cal D}}^{+,w_4}_{j_4}
\right\rangle\, .\label{4d}
\end{equation}
The OPE ${\widehat{\cal D}}^{+,w_1}_{j_1}\otimes{\widehat{\cal
 D}}^{+,w_2}_{j_2}$
 computed
in the previous section (and similarly for $j_3,j_4$) requires
either
$w_1+w_2=-w_3-w_4-1$ or $w_1+w_2=-w_3-w_4- 2$ or
$w_1+w_2=-w_3-w_4-3$ for discrete intermediate states and
 $w_1+w_2=-w_3-w_4- 2$
for continuous intermediate states. And similarly in the other channels.

Repeating this analysis for four-point functions involving fields in
different representations, it is straightforward to conclude that the spectral
flow selection rules  for four-point functions in different sectors
can be obtained from those
 for two- and three-point functions, or equivalently from the OPE found
in section 4.

\section {Summary and conclusions}

We have studied the OPE in the AdS$_3$ WZNW
model. Performing the analytic continuation of the expressions in the
Euclidean $H_3^+$ WZNW model proposed in \cite{tesch1, tesch2} and adding
spectral flow, $i.e.$ considering the full set of structure constants, we
obtained the OPE of spectral flow images of primary fields in the
Lorentzian theory. Assuming the results  also hold for affine
descendants, we have argued that a truncation 
is necessary in order to avoid contradictions and we have shown
that a consistent cut amounts to the closure of the operator algebra
on the Hilbert space of the theory.
Indeed, the spectral flow symmetry  implies that
only operators outside the physical spectrum must be discarded  and
moreover, every physical state contributing to a given OPE is also
found to appear in  all possible equivalent operator products.
The fusion rules obtained in this way are consistent with  results in
\cite{mo3}, deduced from the factorization of four-point functions of
$w=0$ short strings in the boundary conformal field theory, 
and contain in addition operator products
involving states in continuous representations. A discussion of the
relation between our results and some conclusions in \cite{mo3}
can be found in the appendix A.2. 

Implementing the truncation in the procedure 
followed in section 3 in order to directly obtain a consistent OPE 
does not seem possible because it would break
analyticity. Therefore, an inevitable conclusion is that either the
prescription must be modified in order to avoid inconsistencies with the
spectral flow symmetry, $i.e.$ the route we have followed to relate the
OPE in the H$_3^+$ and the AdS$_3$ models is not
self-consistent, or the structure constants must be further constrained.
Nevertheless, although the physical process
determining the truncation is not completely understood, 
several consistency checks have been performed in section 4 and
 the OPE  displayed in  items
1. to 3. can be taken to stand on solid foundations.

The full consistency of the fusion rules should follow from a proof of
factorization and crossing symmetry of the four-point functions.
A preliminary analysis of the factorization of 
amplitudes involving states in different sectors
of the theory was presented in section 5. Based on the factorization
$ansatz$, we proposed an expression for generic four-point functions
and we showed that some terms are redundant in $w-$conserving
amplitudes. 
We illustrated in one example
 that the amplitudes must factorize as expected in order to avoid
 inconsistencies, $i.e.$ if the bootstrap approach
 holds, only states
according to the fusion rules determined in section 4 must
propagate in the intermediate channels. Analogously as the OPE, 
the factorization also agrees with
the spectral flow selection rules.
However more work is necessary to put
 this {\it ansatz} on a firmer mathematical ground. In particular,
 additional information on the action of the spectral flow operation
 on descendant operators is
 required to verify crossing symmetry. 

Given that scattering amplitudes of string
theory on AdS$_3$ should be obtained from
correlation
functions in the ${SL(2,\mathbb R)}$ WZNW model,
our results constitute a step forward towards the construction of the S-matrix in string
theory on Lorentzian AdS$_3$ and
 to learn more about the dual conformal field theory on the
boundary through AdS/CFT, in the spirit of \cite{mo3}. Indeed
an important application of our results would be to construct the
S-matrix of long strings in AdS$_3$ which
describes scatterings in the CFT defined on the 
Lorentzian two-dimensional boundary.
In particular, the OPE
$\widehat{\cal C}_{j_1}^{\alpha_1,w_1} 
\otimes\widehat{\cal C}_{j_2}^{\alpha_2,w_2}$ obtained  in section 4
sustains the expectations in 
\cite{mo3} that short and long strings should appear as poles
in the scattering of asymptotic states of long strings.

\bigskip

\noindent{\bf Acknowledgments:}

We gratefully acknowledge discussions
with G. Aldazabal, E. Andr\'es, C. Cardona Giraldo, S. Iguri,
J. Maldacena and P. Minces and correspondence with S. Ribault and Y. Satoh.
This work was supported in part by PIP 6332
CONICET and by UBACYT X103.

\appendix
\section{Appendices}
\subsection{Analytic structure of $W_1$}

\renewcommand{\theequation}{A-\arabic{equation}}
\setcounter{equation}{0}

The purpose of this appendix is to study the analytic structure of
 $W_1$.
 In particular, we are specially
interested in possible
 zeros appearing in $W_1$ which are not evident in the expression
(\ref{w1}),
 but are very important in our definition of the OPE.

Let us recall some useful identities relating
different expressions for $G\left[\begin{matrix} a,b,c\cr
e,f\cr\end{matrix}\right]$ \cite{slater},
\bea
G\left[\begin{matrix} a,b,c\cr e,f\cr\end{matrix}\right]=
\frac{\Gamma(b)\Gamma(c)}{\Gamma(e-a)\Gamma(f-a)}
G\left[\begin{matrix} e-a,f-a,u\cr
u+b,u+c\cr\end{matrix}\right],\label{idG1}
\eea
\bea
G\left[\begin{matrix} a,b,c\cr e,f\cr\end{matrix}\right]=
\frac{\Gamma(b)\Gamma(c)\Gamma(u)}{\Gamma(f-a)\Gamma(e-b)\Gamma(e-c)}
G\left[\begin{matrix} a,e-b,e-c\cr
e,a+u\cr\end{matrix}\right]\, ,\label{idG2}
\eea
where $u$ is defined
as $u=e+f-a-b-c$. Using the permutation symmetry among $a,b,c$ and
$e,f$, which is evident from the series representation of the
hypergeometric function ${}_3F_2$, seven new
identities may be generated. In what follows we use these
identities in order to obtain the greatest possible amount of
information on $W_1$. 

Consider for instance $C^{12}$
defined in (\ref{Wfull}). Using
(\ref{idG1}), it can be rewritten for $j_1=-m_1+n_1$, with $n_1$ a non
negative 
integer, as
\bea
C^{12}&=&\frac{\Gamma(-N)\Gamma(-j_{13})\Gamma(-j_{12})\Gamma(1+j_2+m_2)}
{\Gamma(-j_3-m_3)}\cr
&\times&\sum_{n=0}^{n_1} \left(\begin{matrix}
n_1\cr n\cr\end{matrix}\right) \frac{(-)^n}{\Gamma(n-2j_1)}
\frac{\Gamma(n-j_{12})}{\Gamma(-j_{12})}\frac{\Gamma(n+1+j_{23})}{\Gamma(1+j_{23})}
\frac{\Gamma(1+j_3-m_3)}{\Gamma(1+j_3-m_3-n_1+n)}.
\label{c12-1}
\eea

Using (\ref{idG2}) instead of (\ref{idG1}), one finds an expression
for $C^{12}$ equal to (\ref{c12-1}) with  $j_3\rightarrow-1-j_3$.

There is a third expression in which 
$C^{12}$ can be written as a finite sum for generic $j_2,j_3$. This follows
from (\ref{Wfull}), using the identity obtained from (\ref{idG2})
with $(e\leftrightarrow f)$. This expression is explicitly
invariant under $j_3\rightarrow-1-j_3$.

Consider for instance 
(\ref{c12-1}). All quotients inside the sum are such that the arguments
in the $\Gamma-$functions of the denominator
equal those in the numerator up to a positive integer,
except for the one with
$\Gamma(n-2j_1)$ which is regular and non vanishing for ${\rm
Re}\,j_1<-\frac 12$.
Then, each
quotient is separately regular. Eventually, some
of them may vanish, but not for all values of $n$. In particular, for
$n=0$ the first two quotients equal one. The last factor may vanish for
$n=0$, but for $n=n_1$ it equals  one.
However, particular configurations of $j_i$, $m_i$
may occur such that one of
the first two quotients vanishes for certain  values of $n$,
namely $n=n_{min}, n_{min}+1,\dots,n_1$, and the last one vanishes
 for other special values, namely $n=0,1,\dots,n_{max}$.
Thus, if $n_{max}\geq n_{min}$, all terms in the sum cancel and
$C^{12}$ vanishes as a simple zero.
   In fact, let us consider for instance  both $1+j_{23}=-p_3$ and
   $1+j_3-m_3=1+n_3$,
with $p_3,n_3$  non negative integers. This requires
$\Phi^{j_2,w_2}_{m_2,\overline m_2}\in {\cal D}^{-,w_2}_{j_2}$
and $j_3=j_1-j_2-1-p_3=m_3+n_1-n_2-1-p_3$, which impose $p_3<n_1$ and
allow to rewrite
the sum in (\ref{c12-1}) as
\bea
\sum_{n=0}^{p_3} \frac{1}{n!}\frac{n_1!}{(n_1-n)!}\frac{p_3!}{(p_3-n)!}
\frac{\Gamma(n-j_{12})}{\Gamma(-j_{12})}\frac{1}{\Gamma(n-2j_1)}
\frac{n_3!}
{\Gamma(1+n_3-n_1+n)}\, .\label{c4}
\eea

Finally, taking into account that $1+n_3-n_1+n=-n_2-(p_3-n)\leq0$,
for $n=0,1,\dots,p_3$, the sum vanishes as a simple zero.
A similar analysis for $j_{12}=p_3\geq0$ and
$1+j_3-m_3=1+n_3\geq1$ shows that
no zeros appear in this case when $\Phi^{j_2,w_2}_{m_2,\overline m_2}$
is the spectral flow image of a primary field.

From the expression obtained for $C^{12}$ by changing $j_3\rightarrow-1-j_3$, 
one finds zeros again for
$\Phi^{j_2,w_2}_{m_2,\overline m_2}\in {\cal D}^{-,w_2}_{j_2}$. These  appear
when both
$j_{3}=j_2-j_1+p_3$ and $j_3=-m_3-1-n_3$ hold simultaneously.

Finally, repeating the analysis for the sum in the third expression
for $C^{12}$, $i.e.$ that explicitly symmetric under $j_3\rightarrow
-1-j_3$,
 one finds the same zeros as in the
previous cases.

Let us now consider the analytic structure of $W_1:=D_1C^{12}\overline
C{}^{12}$.
Expression (\ref{w1}) together with the  discussions above allow to rewrite  $W_1$ as
\bea
W_1(j_i;m_i,\overline m_i )=
\frac{(-)^{m_3-\overline m_3+\overline n_1}\pi^2 \gamma(-N)
}{\gamma(-2j_1)\gamma(1+j_{12})\gamma(1+j_{13})}\frac{\Gamma(1+j_2+m_2)}
{\Gamma(-j_2-\overline m_2)}
\frac{\Gamma(1+j_3+m_3)}{\Gamma(-j_3-\overline m_3)}
E_{12}\overline E{}_{12}\, ,
\eea
where
$E_{12}$ is given by $\Gamma(-2j_1)$ times (\ref{c4}).
$E_{12}$ has
no poles but it may vanish for certain
special configurations if
$\Phi^{j_2,w_2}_{m_2,\overline m_2}\in {\cal D}^{-, w_2}_{j_2}$,
namely $n_2<n_1-p_3$ and $j_3=m_3+n_3$ or
$j_3=-m_3-1-n_3$, with $n_3=0,1,2,\dots$, where $p_3=-1-j_{23}$ 
in the former and $p_3=j_{13}$ in the latter. The same result 
applies to  $\overline E{}_{12}$,  changing
$n_i$ by $\overline n_i$. Obviously one might find, using
other identities, new zeros for special configurations. This
could be a difficult task, because the series does not 
reduce to a finite sum in general. Fortunately,
it is not necessary for our purposes.

\subsection{ Relation to \cite{mo3}}
\renewcommand{\theequation}{B-\arabic{equation}}
\setcounter{equation}{0}

This appendix contains some comments about the relation
between our work and \cite{mo3}. For simplicity, we use the
conventions of the latter,
related  to ours by $j\rightarrow-j$ in the $x-$basis, up to normalizations. 
The range of $j$ for discrete representations is now $\frac 12<j<\frac{k-1}{2}$
and for continuous representations, $j=\frac 12+i\mathbb R$.

One of the aims of \cite{mo3} was
to study the factorization of four-point functions involving $w=0$ short
strings in the boundary conformal field theory.
The $x-$basis seems  
appropriate for this purpose  since $x_i,\overline x_i$ 
can be interpreted as the coordinates of the boundary.
Naturally, both the OPE and the factorization
 look very different in the $m-$ and $x-$basis. 
For instance, it is not obvious how
discrete series would 
appear in the
OPE or factorization of fields
in continuous representations if they are to be obtained from the
analogous expressions in the H$_3^+$ model in the $x-$basis.
However, when discrete representations are involved, there are
certain similarities. Actually,
in accord with the fusion rules
  $\hat{\cal D}_{j_1}^{+,w_1}\otimes\hat{\cal D}_{j_2}^{+,w_2}$ 
obtained
  in section 4,
 $w=1$ long
  strings and $w=0$ short strings were found in the 
  factorization studied in \cite{mo3}. 
Conversely, 
it was interpreted that $w=1$ short strings  do not
  propagate in the intermediate channels, while we found spectral flow
  non-preserving contributions of discrete representations in the OPE.
In this appendix
 we analyze this issue. We reexamine 
the three-point
functions involving two $w=0$ strings and one $w=1$ short string 
and certain divergences in the  four-point functions
of $w=0$ short strings, namely the so-called
Poles$_2$, which seem to break the factorization.

\begin{itemize}
    \item {\it Three-point functions involving one $w=1$ short string and
    two $w=0$ strings}
\end{itemize}

The $w-$conserving two-point functions 
of short strings in the target space ($w\ge 0$) are given by
\bea
\langle\Phi^{\omega,j}_{J,\bar J}(x_1, \overline x_1)\Phi^{\omega,j}_{J,\bar
J}(x_2, \overline x_2)\rangle\sim
|2j-1\pm(k-2)\omega|\frac{\Gamma(2j+p)\Gamma(2j+\bar
p)}{\Gamma(2j)^2p!\bar p!}\frac{{\cal B}(j)}{x_{12}^{2J}\bar
x_{12}^{2\bar J}}\,,\label{tpf}
\eea
where ${\cal B}(j)=B(-j)$ and the upper (lower) sign holds for $J=j+p+\frac k2w$ 
($J=-j-p+\frac k2w$), $p, \overline p$ being non-negative integers. 
Three-point functions of $w=0$ string states are 
\bea
\left\langle \Phi_{j_1}(x_1, \overline x_1) \Phi_{j_2}(x_2, \overline x_2)
\Phi_{j_3}(x_3, \overline x_3)\right\rangle= C(j_1,j_2,j_3)
\prod_{i>j}|x_{ij}|^{-2j_{ij}}\, ,\label{3ptxw=0}
\eea
and for one $w=1$  short string
and two $w=0$ strings they are given by
(we omit the $x, \overline x-$dependence)
\bea
\langle\Phi^{j_1,\omega=1}_{J_1,\bar
J_1}(x_1, \overline x_1)\Phi_{j_2}(x_2, \overline x_2)
\Phi_{j_3}(x_3, \overline x_3)\rangle\sim\frac{1}{\Gamma(0)}{\cal B}(j_1)
C\left(\frac k2-j_1,j_2,j_3\right)\times~~~~~~~~~~~~~\cr 
~~~~~~~~~~~~~~~~~\frac{\Gamma(j_2+j_3-J_1)}{\Gamma(1-j_2-j_3+\bar J_1)}
\frac{\Gamma(j_1+J_1-\frac k2)}{\Gamma(1-j_1-\bar J_1+\frac k2)}
\frac{1}{\gamma(j_1+j_2+j_3-\frac k2)}\, .\label{3ptx}
\eea
The $\Gamma(0)^{-1}$ factor is absent when the $w=1$ operator is a
long string state. This three-point function was obtained in
\cite{mo3} from an equivalent expression in the $m-$basis.  $J_1,
\overline J_1$  label the  global
$SL(2,\mathbb R)$ representations and can be written in terms of 
parameters $m_1, \overline m_1$ as
$J_1=\mp m_1+\frac k2$,
$\overline J_1=\mp\overline m_1+\frac k2$, depending if the correlator
involved the field $\Phi_{m_1,\overline m_1}^{j_1, w_1=\mp 1}$.

As observed in \cite{mo3}, when $J_1=\frac
k2-j_1-p$, $\overline J_1=\frac k2-j_1-\overline p$, 
the factor $\frac{\Gamma(j_1+J_1-\frac
k2)}{\Gamma(1-j_1-\overline J_1+\frac k2)}$ cancels the $\Gamma(0)$ and
the three-point function is finite and
 can be  interpreted as a $w-$conserving amplitude. 
To see this, recall that if it was obtained from a $w=-1$
three-point function in the $m-$basis and
$m_1=j_1+p$, then
\bea
\langle\Phi^{j_1,w=1}_{J_1,\bar
J_1}(x_1,\overline x_1)\Phi_{j_2}(x_2,\overline x_2)\Phi_{j_3}(x_3,
\overline x_3)\rangle\sim  (-)^{p+\bar
p} {\cal B}(j_1)C\left(\frac k2-j_1,j_2,j_3\right)\times~~~~~~~~~~~~~~~~~\cr 
\frac{\Gamma(j_2+j_3+j_1-\frac k2+p)}{p!
\Gamma(j_2+j_3+j_1-\frac k2)}\frac{\Gamma(j_2+j_3+j_1-\frac
k2+\bar p)}{\bar p! \Gamma(j_2+j_3+j_1-\frac k2)}\label{3ptw1}
\eea
 reduces to (\ref{3ptxw=0}) when
$p=\overline p=0$ and $j_1\rightarrow \frac k2-j_1$, as expected from
  spectral flow symmetry. Similarly, if $w=+1$ and $m_1=-j_1-p$, the same
 interpretation holds.

On the contrary, for  $w=-1$ ($w=+1$)
and $m_1=-j_1-p$ ($m_1=j_1+p$),  the $\Gamma(j_1+J_1-\frac k2)$ does not cancel
the factor $\Gamma(0)^{-1}$ and then, it was
concluded in \cite{mo3} that the three-point function 
 vanishes in this case.

However, notice that if
$J_1=\frac
k2+j_1+n=j_2+j_3+p$, $\overline J_1=\frac k2+j_1+\overline
n=j_2+j_3+\overline p$, $n,\overline n\in {\mathbb Z}_{\ge0}$, 
the r.h.s. of (\ref{3ptx}) can also be rewritten as
the r.h.s. of (\ref{3ptw1}), but now
this non-vanishing amplitude corresponds to
a $w=1$ three-point function which is not equivalent to a $w-$conserving one.
Indeed,  (\ref{3ptw1}) is regular as long as $n<p$ ($\overline
n<\overline p$) and when
$n\ge p$ ($\overline n\ge \overline p$)  there are 
divergences in $C(\frac k2-j_1,j_2,j_3)$  at $j_1=j_2+j_3-\frac k2-q$ 
with $q=0,1,2,\cdots$.
Using the spectral flow symmetry, the $w=1$ short string
can be identified with a $w=2$ short string with  $\tilde j_1=\frac
k2-j_1=k-j_2-j_3+q$, which correspond to the
Poles$_2$ in \cite{mo3}.

\begin{itemize}
    \item {\it Factorization of four-point functions of
    $w=0$ short strings}
\end{itemize}

The four-point amplitude of $w=0$ short strings
was  extensively studied in \cite{mo3}. 
The conformal blocks were rearranged  as sums of products of
positive powers of $x$ times  functions of $u=z/x$. In order to perform the
integral over the worldsheet before the $j-$integral, it was
necessary to change the $j-$integration contour from $\frac 12+i\mathbb R$ to
$\frac{k-1}{2}+i\mathbb R$, and
in this process two types of sequences of poles
were picked up, namely \bea &&{\rm Poles}_1:~~j_3=j_1+j_2+n,\cr
&&{\rm Poles}_2:~~j_3=k-j_1-j_2+n, \nonumber\eea where $n=0,1,2,\dots$.
Only values of $n$ for which $j_3<\frac{k-1}{2}$
 contribute to the factorization, so Poles$_1$ appear when
$j_1+j_2<\frac{k-1}{2}$ and Poles$_2$ when
$j_1+j_2>\frac{k+1}{2}$. The contributions from
Poles$_1$ were identified as two
particle states of short strings in the boundary conformal field
theory,
but no interpretation was found
for Poles$_2$ as $s-$channel exchange. 

Recall that we found Poles$_1$ among
the $w-$conserving discrete contributions to the OPE  
$\mathcal
D^{+,w_i}_{j_i}\times \mathcal
D^{+,w_i}_{j_i}$ (see (\ref{1}))
 and Poles$_2$ in the $w-$violating terms with
$\tilde j_3=\frac k2-j_3=j_1+j_2-\frac k2 -n$.
Therefore, it seems tempting to consider
  Poles$_2$ as
 two particle states of $w=1$ short strings in the boundary conformal
 field theory. However, neither the powers of $x, \overline x$ nor the
 residues of the poles in the four-point function studied in
 \cite{mo3} allow this interpretation and thus the Poles$_2$ had to be
 truncated. Clearly,  more work is necessary to determine the
 four-point function and understand
 the factorization.


\subsection{ KZ equations
in the $m-$basis and the factorization {\it ansatz}}
\renewcommand{\theequation}{C-\arabic{equation}}
\setcounter{equation}{0}

We studied some features of the factorization
of four-point functions in section 5.
The purpose of this appendix is to show
some consistency conditions of the expressions used in that section.

Let us start by considering the KZ equation 
for $w-$conserving $n-$point functions in the $m-$basis, namely \cite{ribault}
\bea
\mathcal{E}_i~\kappa^{-1}\left\langle \prod_{\ell=1}^n
\Phi_{m_\ell,\overline{m}_\ell}^{j_\ell,w_\ell}(z_\ell,\overline{z}_\ell)
\right\rangle=0,\label{KZm}
\eea
where
\bea
\mathcal E_i\equiv (k-2)\frac{\partial}{\partial z_i}+\sum_{j\neq i}
\frac{Q_{ij}}{z_{ji}},~~~~Q_{ij}=-2t_i^3t_j^3+t_i^-t_j^++t_i^+t_j^-,
\eea
$t^a$ are defined by $\tilde J_0^a|j,m, \overline m, w>=
-t^a|j,m,\overline m,w>$, $|j,m,\overline m, w>$ 
being the state corresponding to the field
$\Phi_{m,\overline m}^{j,w}$ and $\kappa$ was introduced in section 5. 

Since  a generic
$w-$conserving four-point function can be obtained from the expression
involving four $w=0$ fields,  we concentrate on
\bea
\left\langle \prod_{i=1}^4\Phi_{m_i,\overline{m}_i}^{j_i,w_i=0}
(z_i,\overline{z}_i)\right\rangle&=&|z_{34}|^{2(\tilde\Delta_2+\tilde\Delta_1-
\tilde\Delta_4-\tilde\Delta_3)}|z_{14}|^{2(\tilde\Delta_2+\tilde\Delta_3-
\tilde
\Delta_4-\tilde\Delta_1)}
|z_{13}|^{2(\tilde\Delta_4-\tilde\Delta_1-\tilde\Delta_2-\tilde\Delta_3)}
\nonumber\\
&&\times~ |z_{24}|^{-4\tilde
\Delta_2}~\mathcal F_j(z,\overline z)\, ,\nonumber
\eea
$\mathcal F_j(z,\overline z)$ being a function of the cross ratios $z,
\overline z$, not
determined by conformal symmetry. The KZ equation (\ref{KZm})
implies the following constraint
\bea
\frac{\partial\mathcal
F_j(z,\overline z)}{\partial z}=\frac{1}{k-2}
\left[\frac{Q_{21}}{z}+\frac{Q_{23}}{z-1}\right] \mathcal
F_j(z,\overline z)\, .\label{KZF}
\eea

Assuming that $\mathcal F_j(z,\overline z)$ has the following form
\bea
\mathcal F_j(z,\overline z)=\sum_{N,\overline
N=0}^\infty\int dj \left\{A_j^{(N,\overline N)}\left[\begin{matrix}
j_1\,,\,j_2\,,\,j_3\,,\,j_4\cr m_1,m_2,\dots,\overline
m_4\cr\end{matrix}\right]
z^{\Delta_j-\tilde\Delta_1-\tilde\Delta_2+N}\overline
z^{\Delta_j-\tilde\Delta_1-\tilde\Delta_2+ \overline
N}\right\}\, ,\label{factansm}
\eea
inserting it into
(\ref{KZF}) with
$\Delta_j=\tilde\Delta_j\equiv-\frac{j(1+j)}{k-2}$, then
$A_j^{(0,0)}\left[\begin{matrix}
j_1\,,\,j_2\,,\,j_3\,,\,j_4\cr m_1,m_2,\dots,\overline
m_4\cr\end{matrix}\right]$ satisfies
\bea
&&\left\{2m_1m_2-j(1+j)+j_1(1+j_1)+j_2(1+j_2)\right\}
A_j^{(0,0)}\left[\begin{matrix} j_1\,,\,j_2\,,\,j_3\,,\,j_4\cr
m_1,m_2,\dots,\overline m_4\cr\end{matrix}\right]=~~~~~~~~~~~~~~~~~~~~\nonumber\\
&&~~~~~~~~~~~~~~~~~~~~~~~~~~~~~~~=~(m_1-j_1)(m_2+j_2)A_j^{(0,0)}\left[\begin{matrix}
j_1~,~j_2~,~j_3~,~j_4\cr m_1+1,m_2-1,\dots,\overline
m_4\cr\end{matrix}\right]\nonumber\\
&&~~~~~~~~~~~~~~~~~~~~~~~~~~~~~~~+~(m_1+j_1)(m_2-j_2)A_j^{(0,0)}\left[\begin{matrix}
j_1~,~j_2~,~j_3~,~j_4\cr m_1-1,m_2+1,\dots,\overline
m_4\cr\end{matrix}\right]\, .\label{KZA}
\eea

The equations relating coefficients  
$A_j^{(N,\overline N)}$ with $N, \overline N\ne 0$,
are much more
complicated because they mix terms with different values of $m_i,
\overline m_i$ with terms at different levels $N, \overline N$.

This equation does not have enough information to determine
$A_j^{(0,0)}$ completely. So we just check that
the expression found in (\ref{4ptm}) is consistent with an analysis
performed directly in the
$m-$basis. Inserting 
$A_j^{(0,0)}
\left[\begin{matrix} j_1\,,\,j_2\,,\,j_3\,,\,j_4\cr
m_1,m_2,\dots,\overline m_4\cr\end{matrix}\right]= {\mathbb
A}_j^{w=0}(j_1,\dots,j_4;m_1,\dots,\overline m_4)$ 
into (\ref{KZA}) reproduces the same equation
 with $A_j^{(0,0)}$ replaced by $W(j_1,j_2,j;m_1,m_2,m)$.
Because of the complicated expressions known for $W$, we  focus
on the case in which one of the fields in the four-point function
is a discrete primary, namely $\Phi_{m_1,\overline
m_1}^{j_1,w_1=0}\in\mathcal D_{j_1}^{+,w=0}$. In this case, using
(\ref{w1}) one
can show that
(\ref{KZA}) is equivalent to 
\bea
0&=&\sum_{n=0}^{n_1-1}(-)^n\left(\begin{matrix} n_1\cr
n\cr\end{matrix}\right)\left[j-m+\frac{(m_1-j_1)(1+j_1+m_1)}{n_1+1-n}
+\frac{(m_2-j_2)(1+j_2+m_2)(n_1-n)} {n+1+j+m-n_1}\right]\cr
&&~~~\times \frac{\Gamma(n-j_{1}-j_2+j)}{\Gamma(-j_{1}-j_2+j)}
\frac{\Gamma(n+1+j+j_{2}-j_1)}{\Gamma(1+j+j_{2}-j_1)}
\frac{\Gamma(-2j_1)}{\Gamma(n-2j_1)}
\frac{\Gamma(1+j+m)}{\Gamma(n-n_1+1+j+m)}\cr
~~&-&(-)^{n_1}\left[m_1(1-m_1)+j_1(1+j_1)\right]
\frac{\Gamma(n_1-j_{1}-j_2+j)}{\Gamma(-j_{1}-j_2+j)}
\frac{\Gamma(n_1+1+j+j_{2}-j_1)}{\Gamma(1+j+j_{2}-j_1)}
\frac{\Gamma(-2j_1)}{\Gamma(n_1-2j_1)} \nonumber
\eea 
where $n_1=m_1+j_1$ and
$m=m_1+m_2$.
Using $m-$conservation this can be rewritten as 

\bea
0&=&\sum_{n=0}^{n_1-1}(-)^n\left(\begin{matrix} n_1\cr
n\cr\end{matrix}\right)\left[-n\frac{1-n+2j_1}{n_1+1-n}
+\frac{(n-j_{1}-j_2+j)(n+1+j_2+j-j_1)} {n+1+j+m-n_1}\right]\cr\cr
&& ~~~~~\times~ \frac{\Gamma(n-j_{1}-j_2+j)}{\Gamma(-j_{1}-j_2+j)}
\frac{\Gamma(n+1+j_{2}+j-j_1)}{\Gamma(1+j_{2}+j-j_1)}
\frac{\Gamma(-2j_1)}{\Gamma(n-2j_1)}
\frac{\Gamma(1+j+m)}{\Gamma(n-n_1+1+j+m)}\cr\cr
&-& (-)^{n_1}\left[m_1(1-m_1)+j_1(1+j_1)\right]
\frac{\Gamma(n_1-j_{1}-j_2+j)}{\Gamma(-j_{1}-j_2+j)}
\frac{\Gamma(n_1+1+j_{2}+j-j_1)}{\Gamma(1+j_{2}+j-j_1)}
\frac{\Gamma(-2j_1)}{\Gamma(n_1-2j_1)}\, .\nonumber
\eea 
To see
that this vanishes, it is sufficient to note that 
\bea
&&\sum_{n=0}^{n_1-1}(-)^n\left(\begin{matrix} n_1\cr
n\cr\end{matrix}\right)\left[-n\frac{1-n+2j_1}{n_1+1-n}
\right]\frac{\Gamma(n-j_{1}-j_2+j)}{\Gamma(-j_{1}-j_2+j)}
\frac{\Gamma(n+1+j_{2}+j-j_1)}{\Gamma(1+j_{2}+j-j_1)}
\frac{\Gamma(-2j_1)}{\Gamma(n-2j_1)}\nonumber\\
&&~~~~~~~~~~~~~~~~~\times~\frac{\Gamma(1+j+m)}{\Gamma(n-n_1+1+j+m)}\nonumber\\
&&-(-)^{n_1}\left[m_1(1-m_1)+j_1(1+j_1)\right]
\frac{\Gamma(n_1-j_{1}-j_2+j)}{\Gamma(-j_{1}-j_2+j)}
\frac{\Gamma(n_1+1+j_{2}+j-j_1)}{\Gamma(1+j_{2}+j-j_1)}\frac{\Gamma(-2j_1)}
{\Gamma(n_1-2j_1)}\nonumber\\
&&=-\sum_{\tilde n=0}^{n_1-1}(-)^{\tilde n}
\left(\begin{matrix} n_1\cr
\tilde n\cr\end{matrix}\right)\left[\frac{(\tilde n-j_{1}-j_2+j)
(\tilde n+1+j_{2}+j-j_1)}{\tilde n+1+j+m-n_1}\right]
\frac{\Gamma(\tilde n-j_{1}-j_2+j)}{\Gamma(-j_{1}-j_2+j)}
\nonumber\\
&&~~~~~~~~~~~~~~~~~\times~~\frac{\Gamma(\tilde n+1+j_{2}+j-j_1)}{\Gamma(1+j_{2}+j-j_1)}
\frac{\Gamma(-2j_1)}{\Gamma(\tilde n-2j_1)}
\frac{\Gamma(1+j+m)}{\Gamma(\tilde n-n_1+1+j+m)}\, ,
\nonumber
\eea
where $\tilde n=n-1$.

Let us now discuss the other possible $ansatz$, namely (\ref{Aw1}).
To see that ${\mathbb A}_j^{w=1}$ also verifies the KZ equation,
consider $\Delta_j= 
-\frac{j(1+j)}{k-2}-m-\frac k4 $ and $m=m_1+m_2-\frac k2$ in (\ref{KZF}). 
In this case, the equation to be satisfied by $A_j^{(0,0)}$, obtained
by replacing 
(\ref{factansm}) into
(\ref{KZF}), is the following:
\bea
\left\{2m_1m_2-j(1+j)+j_1(1+j_1)+j_2(1+j_2)-(k-2)(m_1+m_2-\frac k4)\right\}
A_j^{(0,0)}\left[\begin{matrix} j_1\,,\,j_2\,,\,j_3\,,\,j_4\cr
m_1,m_2,\dots,\overline m_4\cr\end{matrix}\right]\nonumber
\eea
\bea
&=&(m_1-j_1)(m_2+j_2)A_j^{(0,0)}\left[\begin{matrix}
j_1~,~j_2~,~j_3~,~j_4\cr m_1+1,m_2-1,\dots,\overline
m_4\cr\end{matrix}\right]\nonumber\\
&&+~ (m_1+j_1)(m_2-j_2)A_j^{(0,0)}\left[\begin{matrix}
j_1~,~j_2~,~j_3~,~j_4\cr m_1-1,m_2+1,\dots,\overline
m_4\cr\end{matrix}\right]\nonumber\\
&&-~(m_2-j_2)(m_3+j_3)A_j^{(0,0)}\left[\begin{matrix}
j_1~,~j_2~,~j_3~,~j_4\cr m_1,m_2+1,m_3-1,\dots,\overline
m_4\cr\end{matrix}\right]\, .
\eea

It is not difficult to check that $A_j^{(0,0)}
\left[\begin{matrix} j_1\,,\,j_2\,,\,j_3\,,\,j_4\cr
m_1,m_2,\dots,\overline m_4\cr\end{matrix}\right]=
{\mathbb A}_j^{w=1}(j_1,\dots,j_4;m_1,\dots,\overline m_4)$ 
is a solution of this equation. 

Obviously, $A_j^{w=-1}$ is also a solution of (\ref{KZF})
 when $\Delta_j= 
-\frac{j(1+j)}{k-2}+m-\frac k4 $ and $m=m_1+m_2+\frac k2$.  

Here, we have considered the simple case of four $w=0$ fields.
However, these results can be generalized for arbitrary  
$w-$conserving correlators using the identity (\ref{kappa}).

\end{document}